\documentclass[%
reprint,
amsmath,amssymb,
aps,
superscriptaddress,
prapplied,
floatfix,
]{revtex4-2}

\usepackage[svgnames,psnames]{xcolor}
\usepackage{url}
\usepackage{graphicx}
\usepackage{comment}
\usepackage{afterpage}
\usepackage{dcolumn}
\usepackage{bm}
\usepackage{subcaption}
\usepackage{booktabs} 
\usepackage{tikz}
\usetikzlibrary{positioning, intersections, angles, quotes}
\usepackage{physics}
\usepackage{mathtools}
\DeclarePairedDelimiter\floor{\lfloor}{\rfloor}
\usepackage[colorlinks,citecolor=DarkGreen,linkcolor=FireBrick,linktocpage,unicode,hypertexnames=false]{hyperref}
\usepackage[capitalize]{cleveref}
\crefname{enumi}{}{}

\begin{document}

\title{Comparative Study of Sampling-based Simulation Costs of Noisy Quantum Circuits}

\author{Shigeo Hakkaku}
\affiliation{%
  Division of Advanced Electronics and Optical Science, Department of Systems Innovation, Graduate School of Engineering Science, Osaka University, 1-3 Machikaneyama, Toyonaka, Osaka 560-8531, Japan
}%
\email{shigeo.hakkaku@qc.ee.es.osaka-u.ac.jp}
\author{Keisuke Fujii}%
\affiliation{%
  Division of Advanced Electronics and Optical Science, Department of Systems Innovation, Graduate School of Engineering Science, Osaka University, 1-3 Machikaneyama, Toyonaka, Osaka 560-8531, Japan
}%
\affiliation{%
  Center for Quantum Information and Quantum Biology,
  Institute for Open and Transdisciplinary Research Initiatives,
  Osaka University, 1-2 Machikaneyama, Toyonaka 560-8531, Japan
}%
\affiliation{%
  RIKEN Center for Quantum Computing (RQC),
  Hirosawa 2-1, Wako, Saitama 351-0198, Japan
}%
\email{fujii@qc.ee.es.osaka-u.ac.jp}
\date{\today}

\begin{abstract}
    Noise in quantum operations often negates the advantage of quantum computation. However, most classical simulations of quantum computers calculate the ideal probability amplitudes either storing full state vectors or using sophisticated tensor network contractions.
  Here, we investigate sampling-based classical simulation methods for noisy quantum circuits.
  Specifically, we characterize the simulation costs of two major schemes, stabilizer-state sampling of magic states and Heisenberg propagation, for quantum circuits being subject to stochastic Pauli noise, such as depolarizing and dephasing noise.
  To this end, we introduce several techniques for the stabilizer-state sampling to reduce the simulation costs under such noise.
  It revealed that in the low noise regime, stabilizer-state sampling results in a smaller sampling cost, while Heisenberg propagation is better in the high noise regime.
  Furthermore, for a high depolarizing noise rate $\sim 10\%$, these methods provide better scaling compared to that given by the low-rank stabilizer decomposition.
  We believe that these knowledge of classical simulation costs is useful to squeeze possible quantum advantage on near-term noisy quantum devices
  as well as efficient classical simulation methods.
\end{abstract}

\maketitle

\section{Introduction}\label{sec:intro}
Quantum computers are expected to provide an exponential speedup compared to classical computers for certain problems such as factoring problems and quantum simulations.
Recently, extensive effort has been expended for the realization of quantum computers, and the number of qubits is now more than 50 ~\cite{neillBlueprintDemonstratingQuantum2018,kandalaErrorMitigationExtends2019,aruteQuantumSupremacyUsing2019,aruteHartreeFockSuperconductingQubit2020,jurcevicDemonstrationQuantumVolume2020}.
Unfortunately, such a number of qubits 
is still too small to
run sophisticated quantum algorithms such as Shor's algorithm for factorization.
Nevertheless, 
quantum devices that have already been realized or will be realized soon.
are thought to achieve quantum computational supremacy,
providing an output of a particular task
much faster than the known best effort on a classical computer.
Recently, Google demonstrated a sampling task on its quantum processor, Sycamore, consisting of 53 qubits with high-fidelity single-qubit and two-qubit gates~\cite{aruteQuantumSupremacyUsing2019}.
While Sycamore takes about 200 s to conduct the task, a state-of-the-art supercomputer would take, according to Google's estimation, approximately 10,000 years.
After this experiment, IBM and Alibaba rebutted the estimated time for a classical simulation~\cite{pednaultLeveragingSecondaryStorage2019,huangClassicalSimulationQuantum2020}.
Specifically, Alibaba estimated that the simulation takes less than 20 days by use of a sophisticated designed tensor network-based classical simulation algorithm~\cite{huangClassicalSimulationQuantum2020}.
While noise on quantum operations
deteriorates the quantumness, which would reduce the classical simulation cost,
the effect of the noise is not fully used
in the above-mentioned classical simulations.
To compare 
quantum and classical computers fairly,
a more-refined classical approach 
should be used to simulate quantum computers.

There has been another classical simulation approach 
based on sampling.
Instead of computing the full probability amplitudes, a class of classically simulatable quantum computations, which is called a Clifford circuit, is used to save memory during a classical simulation~\cite{howardApplicationResourceTheory2017}.
More precisely, a quantum circuit is decomposed into
the Clifford circuits and the preparation of a resource state which is the so-called magic state~\cite{bravyiUniversalQuantumComputation2005}.
Then, by decomposition of the resource state into 
a linear combination of stabilizer states with a quasiprobability distribution, 
the resource state is replaced by sampling a stabilizer state with an appropriate postprocessing,
where the Gottesman-Knill theorem~\cite{aaronsonImprovedSimulationStabilizer2004} can efficiently simulate each realization.
We refer to this sampling algorithm as ``stabilizer-state sampling."
The sampling cost of this quantum computation is determined by a measure called the ``robustness of magic (ROM)."

Another sampling-based classical simulation algorithm is based on evolving a measured observable in the Heisenberg picture~\cite{rallSimulationQubitQuantum2019}.
The observable evolved by an adjoint quantum channel is decomposed over the Pauli operators, and one of them is sampled by the quasiprobability method similarly to stabilizer-state sampling.
This sampling algorithm is called ``Heisenberg propagation."
The simulation costs are characterized by a measure called a ``stabilizer norm."
A study on the simulation cost of a depolarized rotation gate showed that the depolarizing noise decreases the simulation costs~\cite{rallSimulationQubitQuantum2019}.
However, it remains unclear whether Heisenberg propagation can simulate noisy quantum circuits more efficiently than stabilizer-state sampling.

In this study, we investigate the simulation costs of noisy quantum circuits in further detail.
Specifically, we consider two algorithms, stabilizer-state sampling and Heisenberg propagation, to quantify the sampling costs of quantum circuits subject to stochastic Pauli noise.
Unfortunately, 
the existing stabilizer-state sampling has not been optimized to simulate noisy quantum circuits.
Unlike Heisenberg propagation, stabilizer-state sampling is not directly applicable to quantum circuits where nondiagonal noise, such as Pauli $X$ and $Y$ errors, occurs.
Moreover, noise on the Clifford gates has yet to be fully used to reduce simulation costs.
To address the former problem, we use the gate-teleportation technique~\cite{zhouMethodologyQuantumLogic2000} to transform the nondiagonal part to a diagonal noise.
For the latter issue, we propose a method to collect noise on Clifford gates into a resource state.
These techniques reduce the simulation costs for stabilizer-state sampling and allows us to compare the two sampling-based simulation algorithms.
In addition, we introduce a reduced stabilizer basis to calculate a reasonable upper bound of the ROM for multiple copies of a noisy magic state in a feasible way.

For these two major sampling-based classical simulation algorithms, we compare the costs to simulate noisy quantum circuits and identify the more-suitable approach in different situations.
Specifically, we consider noisy quantum circuits where each gate is followed by dephasing or depolarizing noise.
We quantitatively analyze how such noise decreases the simulation costs for the two sampling-based simulation algorithms.
We find that there is a crossover in the performance: up to a particular error rate, stabilizer-state sampling has better performance; however, as the error rate becomes higher, Heisenberg propagation outperforms stabilizer-state sampling.
This knowledge is useful to pursue a better approach to simulate a noisy quantum circuit. 
Furthermore, specified classically simulatable regions would be also helpful to design quantum circuits
that potentially have a quantum advantage avoiding these sampling-based classical simulations.

The rest of this paper is organized as follows.
In Section~\ref{sec:sampling_alg_review}, we review two existing classical simulation algorithms: stabilizer-state sampling and Heisenberg propagation.
The analysis includes the simulation costs, ROM and stabilizer norm.
Then, in Section~\ref{sec:sampling_costs_of_noisy_circuits}, we explain how to calculate the ROM of noisy quantum circuits through some examples.
Finally, in Section~\ref{sec:comparison_two_methods}, we compare the two sampling-based algorithms via noisy random quantum circuits (RQCs).
Section V is devoted to the conclusions and a discussion.

\section{Sampling-based classical simulation}\label{sec:sampling_alg_review}
We initially review two existing classical simulation algorithms, 
stabilizer-state sampling~\cite{howardApplicationResourceTheory2017} and Heisenberg propagation~\cite{rallSimulationQubitQuantum2019}.
They are similar in the sense that 
a state or an operator is decomposed into a linear combination of 
basic states or operators that can be simulated efficiently.
Then one of the basic states or operators is sampled via a quasiprobability method.

\subsection{Stabilizer-state sampling}
\label{subsec:preliminary}
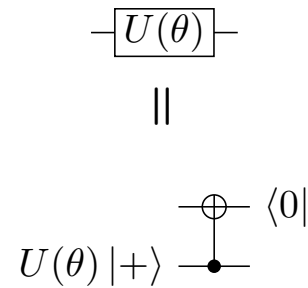
\begin{figure}[tb]
  \begin{subfigure}[t]{\linewidth}
    \centering
    \label{subfig:rotation_gate}
    \includegraphics[scale=1.5]{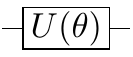}
  \end{subfigure}
  \begin{subfigure}[t]{\linewidth}
    \centering
    \begin{tikzpicture}
      \node[rotate=90, scale=2, font=\bfseries] at (0,0) {=};
    \end{tikzpicture}
  \end{subfigure}
  \begin{subfigure}[t]{\linewidth}
    \centering
    \label{subfig:gadgetized_rotation_gate}
    \includegraphics[scale=1.5]{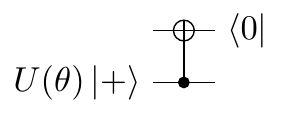}
  \end{subfigure}
  \caption{
      Top: Single-qubit rotation gate about the $z$ axis, $U\pqty{\theta}$.
      Bottom: Gate teleportation of $U\pqty{\theta}$ composed of a controlled NOT gate and measurement in the $Z$ basis with the nonstabilizer resource state $U\pqty{\theta}\ket{+}$.
      The state is postselected on the measurement outcome 0.
      This is justified because we are using teleportation just to translate a quantum circuit for classical simulation.
      In the following figures, projections to $\ket{0}$ are made in the same way.
      }
  \label{fig:rotation_gate_and_gadgetized_rotation_gate}
\end{figure}
Here we explain stabilizer-state sampling on the basis of  Ref.~\cite{howardApplicationResourceTheory2017}.
In general, a quantum circuit 
can be decomposed into Clifford gates and non-Clifford gates.
Clifford gates can be simulated efficiently by
the Gottesman-Knill theorem~\cite{aaronsonImprovedSimulationStabilizer2004}, whereas non-Clifford gates, such as the $T$ gate cannot.
To handle this, we use gate teleportation~\cite{zhouMethodologyQuantumLogic2000}, which replaces a quantum gate with a Clifford circuit and the preparation of a resource state.
More concretely, a single-qubit rotation through angle $\theta$ around the $z$ axis
\begin{align}
  U(\theta) = \op{0} + e^{i\theta}\op{1} \label{eq:rot_gate_around_z}
\end{align}
can be replaced with the preparation of the resource state
\begin{align}
  \ket{U\pqty{\theta}} \coloneqq U\pqty{\theta}\ket{+}, \label{eq:resource_state_of_rot_z}
\end{align}
where
\begin{align}
\ket{+} \coloneqq \frac{1}{\sqrt{2}} \pqty{\ket{0} + \ket{1}},
\end{align}
via gate teleportation as shown in~\Cref{fig:rotation_gate_and_gadgetized_rotation_gate}. The gate teleportation consists of a controlled-Not gate, measurement in the $Z$ basis, and the preparation of the resource state $\ket{U\pqty{\theta}}$.
On the measurement, the state is projected to $\ket{0}$ so that there is no by-product.
This is justified because we are using teleportation just to translate a quantum circuit for classical simulation.
In the following figures, projections to $\ket{0}$ are made in the same way.
When $\theta = \pi/4$, $U(\pi/4)$ is specifically called the ``$T$ gate," and 
its corresponding resource state 
\begin{align}
  \ket{T} \coloneqq U\pqty{\pi/4}\ket{+}
\end{align}
is called the ``magic state"~\cite{bravyiUniversalQuantumComputation2005}.
Because the resource state is not a stabilizer state, it cannot be simulated efficiently. 
To handle this,
we decompose the resource state into a linear combination of stabilizer states
by using the fact that stabilizer states form
an overcomplete basis on the operator space.
In the case of the magic state,
we have 
\begin{align}
  \op{T} =&
  \frac{1+\sqrt{2}}{4} \left(\frac{I+X}{2}  + \frac{I+Y}{2}\right)
  \nonumber \\ 
  +& \frac{1-\sqrt{2}}{4} \left(\frac{I-X}{2}  + \frac{I-Y}{2}\right).
\end{align}
Note that the coefficients are no longer positive
and form a quasiprobability distribution.

Consider 
an arbitrary $n$-qubit universal quantum circuit $U$,
where we want to calculate the expectation value of a Pauli 
operator $A \in \mathcal{P}_n \coloneqq \Bqty{I,X,Y,Z}^{\otimes n}$:
\begin{align}
  \ev{A} = \bra{0}^{\otimes n} U^\dag A U \ket{0}^{\otimes n}.
\end{align}
By decomposing the quantum circuit $U$ into Clifford and $T$ gates,
we can rewrite the expectation value $\ev{A}$
as 
\begin{align}
  \ev{A}
&= 2^{t} {\rm Tr} \left[ \left(A\otimes \op{0}^{\otimes t} \right) U_\text{Cl} 
\left(\op{0}^{\otimes n} \op{T}^{\otimes t} \right) U_\text{Cl}^{\dag} \right]
\end{align}
where $t$ is the number of $T$ gates and $U_\text{Cl}$ is a Clifford circuit (we assume that there is a local description of $U$ so that we can efficiently decompose $U$ into Clifford and $T$ gates.).
The projector $\op{0}^{\otimes t}$ is used for the gate teleportation, 
and $2^{t}$ is due to its normalization.
While we decompose the given quantum circuit into Clifford and $T$ gates for 
simplicity, any diagonal non-Clifford gate can be used straightforwardly instead of using $T$ gates. 

A classical simulation based on stabilizer-state sampling runs as follows:
\begin{enumerate}
  \item Let $\Bqty{\sigma_i}$ be a set of pure $t$-qubit stabilizer states. 
    The product of the magic states $\ket{T}^{\otimes t}$ as a resource state is now decomposed into 
    a linear combination of stabilizer states 
    \begin{align}
      \op{T} ^{\otimes t} = \sum _i x_i \sigma_i,
      \label{eq:decomposition}
    \end{align}
    where $\sum_{i} x_i = 1$.

  \item Sample a stabilizer state $\sigma_i$ with probability 
    \begin{align}
      p_i = \frac{\abs{x_i}}{\sum_i\abs{x_i}}. 
    \end{align}
    \label{enumi:sampling}

  \item Calculate $M_i = \text{sign}(x_i)  \sum_i\abs{x_i} \langle A \rangle _{\sigma_i} $,
    where $\langle A \rangle _{\sigma_i}$ is defined as
    \begin{align}
      \ev{A}_{\sigma_i}  \coloneqq 2^t \Tr\bqty{\pqty{A\otimes \op{0}^{\otimes t}} U_\text{cl} \pqty{\op{0}^{\otimes n} \otimes  \sigma_i} U_\text{cl}^{\dag}},
      \label{eq:A_sigma}
    \end{align} 
    and can be calculated efficiently.
    \label{enumi:calculate_M}

  \item By repeating steps \cref{enumi:sampling} and \cref{enumi:calculate_M}, 
    the expectation value $\text{Ex}(M_i)$ of $M_i$ is estimated.
\end{enumerate}
From \cref{eq:decomposition,eq:A_sigma},
we have 
\begin{align}
  \ev{A}
   &= 2^{t} \sum _i x_i\Tr\bqty{\pqty{A\otimes \op{0}^{\otimes t}} U_\text{cl} 
   \pqty{\op{0}^{\otimes n} \sigma_i} U_\text{cl}^{\dag}} \notag
   \\
   &=  \sum_i p_i \pqty{\sum _j\abs{x_j}} \text{sign}\pqty{x_i} \ev{A}_{\sigma_i} \notag
   \\
   &= \sum_i  p_i M_i .
\end{align}
Therefore, $M_i$ is an unbiased estimator of $\ev{A}$.
Since the sampling according to quasiprobability distribution is simulated by the postprocessing, we call it a ``quasi-probability method."

By use of the Hoeffding inequality~\cite{hoeffdingProbabilityInequalitiesSums1963}, 
the number of samples 
necessary to obtain the expectation value within an additive error $\delta$
with probability of at least $1-\epsilon$
is given by
\begin{align}
  N_\text{stabilizer} = \pqty{\sum_i\abs{x_i}}^2\frac{2}{\delta^2}\ln\pqty{\frac{2}{\epsilon}}.
\end{align}
The sampling cost is proportional to the square of $\sum_i\abs{x_i}$.

Until now, we have considered that the resource state is in a pure state $\op{T}$.
However, in general, the resource state can be in a mixed state possibly due to noise on the $T$ gate.
In the following, 
we simply denote a resource state by $\rho$ in general.
If $\rho$ is a probabilistic mixture of stabilizer states, for example, all the coefficients $x_i$ are positive by definition, and $\sum_i \abs{x_i} = 1$.
Hence, the sampling cost does not increase.
Otherwise, some coefficients are inevitably negative, 
and the sampling cost $N_\text{stabilizer}$ increases according to the amount of $\sum_i\abs{x_i}$.
In this sense, $\sum_i\abs{x_i}$ quantifies the simulation cost of the universal quantum computation.

With use of the minimization 
for all possible stabilizer decompositions,
the ROM of a resource state $\rho$ is defined as follows
\begin{align}
  \mathcal{R}\pqty{\rho} = \min_{x_i}\Bqty{\sum_i\abs{x_i};\rho = \sum_i x_i \sigma_i},
\end{align}
which determines the minimum cost to simulate a Clifford circuit with an input state $\rho$ by stabilizer-state sampling.
However, calculation of the ROM is intractable if the number of qubits increases.
Even in this case, 
we can calculate an upper bound of the ROM by using 
the submultiplicativity,
\begin{align}
  \mathcal{R}\pqty{\rho \otimes \rho'} \leq \mathcal{R}\pqty{\rho} \mathcal{R}\pqty{\rho'}.
\end{align}
Specifically, for the resource state $\op{T}^{\otimes t}$,
it has been estimated that
\begin{align}
  \mathcal{R}(\op{T}^{\otimes 5}) = 3.68705
\end{align}
and hence
\begin{align}
  \Bqty{\mathcal{R} \pqty{\op{T}^{\otimes t}}}^2 < \Bqty{\mathcal{R}\pqty{\op{T}^{\otimes 5}}}^{\frac{2t}{5}} \approx 2^{0.75298 t}
\end{align}
as explained in Ref.~\cite{howardApplicationResourceTheory2017}.
On the other hand, the upper bound of the cost of the resource state $\op{T}^{\otimes t}$ is $2^t$ if we use $\mathcal{R}\pqty{\op{T}}=\sqrt{2}$.
Consequently, optimizing the stabilizer decomposition over multiple copies of a resource state significantly reduces the simulation cost.

\subsection{Heisenberg propagation}
We now review Heisenberg propagation as proposed in Ref.~\cite{rallSimulationQubitQuantum2019}.
In Heisenberg propagation, 
an observable to be measured evolves in the Heisenberg picture.
In general, such a calculation requires exponential time.
To handle this,
we decompose the observable into a linear combination of Pauli operators.
As described below, one Pauli operator is sampled at each step by using quasiprobability method as follows.

Consider an $n$-qubit quantum circuit consisting of $d$ quantum channels $\Bqty{\Lambda_i}_{i=1}^d$.
Each quantum channel acts on a constant number of qubits.
We assume that both the initial state of the circuit $\rho$ and an observable $A$ 
are products of operators acting on a finite number of qubits.
The expectation value of $A$ is written as
\begin{align}
  \ev{A} &= \Tr\pqty{A\Lambda_d \circ \cdots \circ \Lambda_2 \circ \Lambda_1\pqty{\rho}} \notag\\
         &= \Tr\pqty{\rho \Lambda_1^\dag \circ \cdots \circ \Lambda_{d-1}^\dag \circ \Lambda_d^\dag \pqty{A}},
\end{align}
where $\Lambda ^{\dag}$ is the adjoint of $\Lambda$; that is,
\begin{align}
  \Tr\bqty{C \Lambda \qty(B)} = \Tr\bqty{\Lambda^{\dag}\qty(C) B},
\end{align}
for any operators $B$ and $C$.

The Heisenberg propagation algorithm is as follows:
\begin{enumerate}

  \item Decompose $A^{(i)}$ into Pauli operators $\sigma \in \mathcal{P}_n = \Bqty{I,X,Y,Z}^{\otimes n}$:
    \begin{align}
      A^{(i)} = \sum_{\sigma \in\mathcal{P}_n} c_\sigma \sigma,
    \end{align}
    where $ A^{(0)} = A$ and $A^{(i)}$ ($i = 1,\ldots,d$) is defined recursively below.

  \item Sample a Pauli operator $\sigma ^{(i)}$ with probability 
    \begin{align}
      p_{\sigma} = \frac{\abs{c_ {\sigma} }}{\sum_{{\sigma} \in \mathcal{P}_n}\abs{c_{\sigma}}}
      = \frac{\abs{\Tr(A^{(i)} {\sigma})}}{2^n \mathcal{D}\pqty{A^{(i)}}},
    \end{align}
    where $\mathcal{D}\pqty{A} \coloneqq  2^{-n}\sum_\sigma \abs{\Tr(A\sigma)} = \sum_{{\sigma} \in \mathcal{P}_n}\abs{c_{\sigma}} $, which is called the ``stabilizer norm" in Ref.~\cite{campbellCatalysisActivationMagic2011}.
  \item Define $A^{(i+1)} = \Lambda_{d-i}^\dagger\pqty{A^{(i)}}$.    
    \label{enumi:Heisenberg_decompose}

  \item Repeat steps 1-3 for $i = 0,\ldots,d$ to obtain $A^{(d+1)}$, which is an operator still acting on a finite number of qubits. Then we calculate 
    \begin{align}
      M\pqty{\Bqty{\sigma^{(i)}}} \coloneqq \Tr\bqty{A^{(d+1)} \rho } \prod _{i=0}^{d} \bqty{\text{sign}\pqty{c_{\sigma^{(i)}}}\mathcal{D}\pqty{A^{(i)}}}.
    \end{align}

  \item Repeat step 4 to estimate the expectation value of $M\pqty{\Bqty{ \sigma _i }}$.
\end{enumerate}
The expectation value is given by 
\begin{align}
  \sum _{\Bqty{\sigma^{(i)}}} \bqty{\pqty{ \prod _{i=0}^{d} p_{\sigma^{(i)}}} M\pqty{\Bqty{ \sigma^{(i)} }}} = \ev{A},
\end{align}
where $\sum _{\Bqty{\sigma^{(i)}}}$ indicates the summation taken over all possible trajectories.
We used the fact that $\text{sign}\bqty{\Tr(A^{(i)} \sigma^{(i)})}\mathcal{D}\pqty{A^{(i)}}$ is an unbiased estimator for $A^{(i)}$ in step \cref{enumi:Heisenberg_decompose}.
The number of samples required for an additive error $\delta$ with probability of at least $1-\epsilon$  is bounded by the Hoeffding inequality~\cite{hoeffdingProbabilityInequalitiesSums1963}
\begin{align}
  N_\text{Heisenberg} = \pqty{\mathcal{D}\pqty{A} \prod_{i=1}^{d} \mathcal{D}\pqty{\Lambda_i^\dagger}}^2 \frac{2}{\delta^2}\ln\pqty{\frac{2}{\epsilon}},\label{eq:sim_costs_Pauli}
\end{align}
where the channel stabilizer norm for a channel $\Lambda$ is defined by
\begin{align}
  \mathcal{D}\pqty{\Lambda} \coloneqq \max_{\sigma \in \mathcal{P}_n} \mathcal{D}\bqty{\Lambda\pqty{\sigma}}. \label{eq:ch_st_norm}
\end{align}
Since 
\begin{align}
  \mathcal{D} (\Lambda \circ \Lambda' ) \leq \mathcal{D} (\Lambda ) \mathcal{D} (\Lambda' ),
\end{align}
we should choose the channel $\Lambda$ to be simulated appropriately
so that the channel stabilizer norm is minimized while the dimension of $\Lambda$ is maintained tractable.

\subsection{Qualitative comparison of the two algorithms}
The two sampling-based algorithms are similar in the sense that 
if a quantum circuit consists only of Clifford gates,
then the overhead does not grow exponentially.
Both algorithms require an optimization procedure to estimate the simulation costs. That is,
the size of the resource state in stabilizer-state sampling
or the size of the support of the channel must be small
to feasibly calculate the ROM or a channel stabilizer norm , respectively.
For stabilizer-state sampling,
the computationally hard part is imposed by the preparation of a resource state,
which can reduce the simulation cost 
for an arbitrary
quantum circuit by preparing multiple copies of the resource state and decomposing them over stabilizer states.
This could be advantageous against Heisenberg propagation
because its simulation cost explicitly depends on the quantum circuit to be simulated instead of a resource state.

On the other hand,
it is easier for Heisenberg propagation to use the noise effect to reduce the simulation cost.
For example, depolarizing noise, which appears after the Clifford gate,
helps to reduce the overhead in a straightforward manner.
However stabilizer-state sampling requires a special treatment, as developed in the next section.
Because of these complicated factors, it remains unclear which algorithm is better suited for a given noisy quantum circuit.
This is one of the main targets to be clarified in this work.

\section{Stabilizer-state sampling for noisy quantum circuits}
\label{sec:sampling_costs_of_noisy_circuits}
Heisenberg propagation can simulate noisy quantum circuits straightforwardly.
However, it is not straightforward for stabilizer-state sampling to use the noise effect, especially on the Clifford gate, to reduce the simulation cost.
Here we provide several techniques to extend stabilizer-state sampling to the noisy case.

Below, we first consider the simulation cost of a diagonal gate followed by diagonal noise, such as dephasing noise.
Second, we calculate the simulation cost when stochastic Pauli noise occurs
after a non-Clifford gate.
Third, we explain noise fusion, where
the noise occurring in different gates is merged so that 
the noise on the Clifford gates can
reduce the simulation cost of a non-Clifford gate.
Finally, we explain how to reduce the number of stabilizer states as a basis 
to calculate a reasonable upper bound of the ROM of the noisy resource states, since an exact calculation of the ROM is hard.

\subsection{Noise teleportation}
\label{subsec:noise_teleportation}
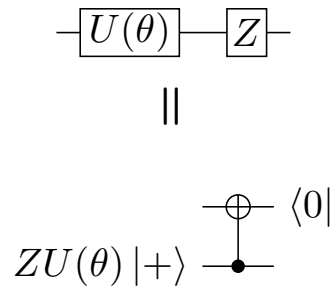
\begin{figure}[tb]
  \centering
  \begin{subfigure}[t]{\linewidth}
    \centering
    \label{subfig:rotation_gate_with_Z_error}
    \includegraphics[scale=1.5]{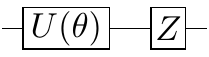}
  \end{subfigure}
  \begin{subfigure}[t]{\linewidth}
    \centering
    \begin{tikzpicture}
      \node[rotate=90, scale=2, font=\bfseries] at (0,0) {=};
    \end{tikzpicture}
  \end{subfigure}
  \begin{subfigure}[t]{\linewidth}
    \centering
    \label{subfig:gadgetized_rotation_gate_with_Z_error}
    \includegraphics[scale=1.5]{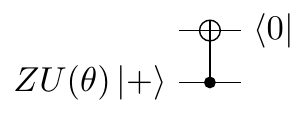}
  \end{subfigure}
  \caption{
     Top: Single-qubit rotation gate about the $z$ axis $U\pqty{\theta}$ followed by the $Z$ error.
    Bottom: Pushing the $Z$ error into the resource state.
  }
  \label{fig:u_theta_with_dephasing_gate_tele}
\end{figure}
\begin{figure}[tb]
  \centering
  \includegraphics[width=\linewidth]{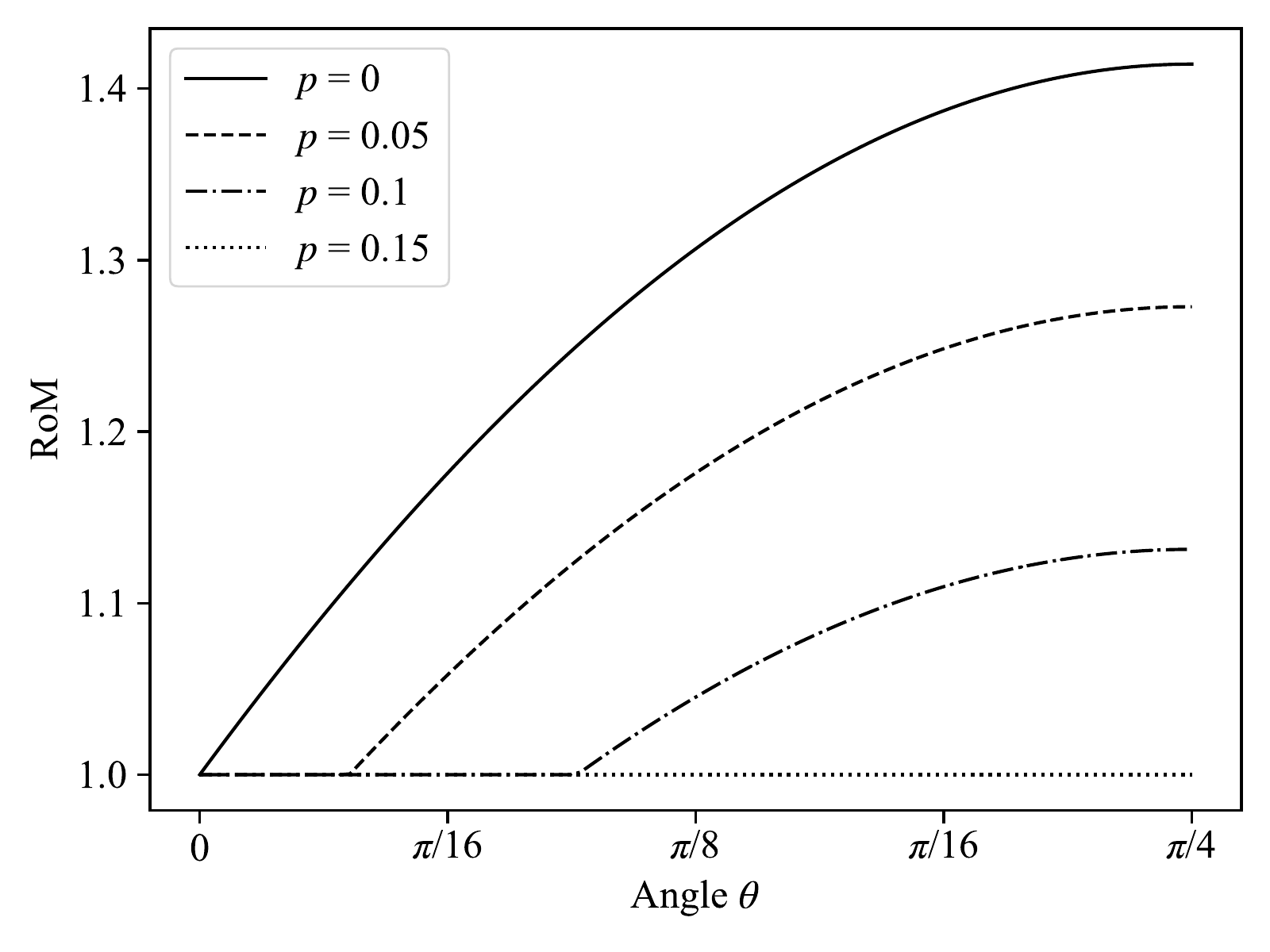}
  \caption{
    ROM of the resource state $\mathcal{E}_\text{dephasing}\bqty{\op{U\pqty{\theta}}}$, which corresponds to the rotation gate about the $z$ axis followed by the dephasing noise with error rate $p$.
    The horizontal axis shows the angle $\theta$ of the rotation.
    The vertical axis shows the ROM, which quantifies the simulation cost of a Clifford circuit with a non-Clifford resource state for stabilizer-state sampling.
    The legend shows the error rates of the dephasing noise.
    The sampling cost increases as ROM becomes larger.
    When the ROM of a circuit is unity, the noisy resource state is a probabilistic mixture of stabilizer states.
  }\label{fig:cost_with_p_and_theta}
\end{figure}
Consider a single-qubit rotation gate about the $z$ axis subject to the dephasing noise.
The ideal rotation gate and the corresponding resource state are defined in \cref{eq:rot_gate_around_z,eq:resource_state_of_rot_z}, respectively.
The single-qubit dephasing noise is defined as
\begin{align}
  \mathcal{E}_\text{dephasing}\pqty{\rho} \coloneqq \pqty{1 - p} \rho + p Z \rho Z. \label{eq:dephasing_noise}
\end{align}
The $Z$ error in \cref{eq:dephasing_noise} acts as the $Z$ operator on the resource state $\ket{U\pqty{\theta}}$, as shown in \cref{fig:u_theta_with_dephasing_gate_tele}.
Thus, the resource state of the noisy rotation gate is given by
\begin{align}
  \rho = \pqty{1-p}\op{U\pqty{\theta}} + p Z\op{U\pqty{\theta}}Z.
\end{align}
\Cref{fig:cost_with_p_and_theta} shows the ROM of the resource state corresponding to the noisy rotation gate  $\mathcal{R}\pqty{\rho}$, where 
we use a convex-optimization solver, CVXPY~\cite{agrawalRewritingSystemConvex2018,diamondCVXPYPythonEmbeddedModeling2016}, to calculate the ROM.
As shown in \cref{fig:cost_with_p_and_theta}, the ROM decreases as the error rate of the dephasing noise increases.
Additionally, the smaller the rotation angle, the more easily the noise makes the ROM unity; that is,
such a noisy resource state becomes a probabilistic mixture of stabilizer states.
One implication of this is that 
we have to carefully design quantum circuits for 
noisy near-term quantum devices so that they cannot be simulated 
easily. 
For example, the variational quantum eigensolver (VQE) uses parameterized quantum circuits consisting of many rotation gates, whose angles are often small.
A small amount of noise would be enough to make such circuits classically simulatable. It would be interesting to characterize applications of noisy near-term quantum devices in terms of the ROM if they satisfy the necessary condition for quantum advantage.

\begin{figure}[tb]
  \begin{subfigure}[t]{\linewidth}
    \centering
    \caption{}
    \label{subfig:diag_U_with_X_error}
    \includegraphics[scale=1.5]{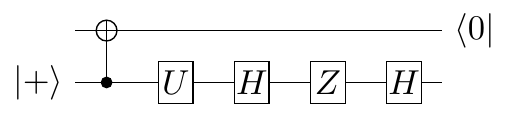}
  \end{subfigure}

  \begin{subfigure}[t]{\linewidth}
    \centering
    \caption{}
    \label{subfig:gadgetized_diag_U_with_X_error}
    \includegraphics[scale=1.5]{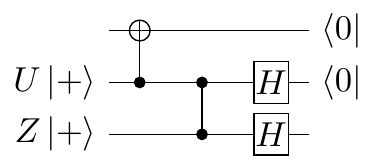}
  \end{subfigure}
  \caption{
    Noise teleportation of a diagonal gate $U$ followed by an $X$ error.
    \subref{subfig:diag_U_with_X_error} Gate teleportation of $U$ followed by an $HZH$ operator.
    \subref{subfig:gadgetized_diag_U_with_X_error} Gate teleportation of $U$ and an $X$ error by replacement of the first $H$ gate shown in \subref{subfig:diag_U_with_X_error} with the gate teleportation of the $H$ gate.
  }
  \label{fig:gadgetized_diag_U_with_X_error}
\end{figure}
In the above case, the noise effect is considered straightforwardly by virtue of the dephasing noise.
However, this argument is not directly applicable for an $X$ error or a $Y$ error because they are not diagonal with respect to the computational basis.
To handle this, we propose a way to deal with nondiagonal errors through gate teleportation.
Consider a diagonal non-Clifford single-qubit gate $U$ followed by the single-qubit depolarizing noise
\begin{align}
  \mathcal{E}_\text{depo1} \pqty{\rho}
  &\coloneqq \pqty{1-\frac{3}{4}p}\rho + \frac{p}{4}\sum_{\mathclap{A\in\Bqty{X,Y,Z}}} \; \bqty{A}\rho \notag\\
  &= \pqty{1-p}\rho + p\frac{I}{2}
\end{align}
where $\bqty{A}$ is a superoperator defined by $\bqty{A} \rho \coloneqq A\rho A^{\dag}$.
A $Z$ error is treated in the same way as mentioned before.
\Cref{subfig:diag_U_with_X_error} corresponds to the gate teleportation of a diagonal gate $U$ followed by an $X$ error,
where the $X$ error is rewritten as $HZH$.
In \Cref{subfig:diag_U_with_X_error} the first $H$ gate is replaced by the gate teleportation of the $H$ gate (\cref{subfig:gadgetized_diag_U_with_X_error}).
As a result, the $X$ error is taken as the $Z$ operator on the second ancilla qubit.
Similarly, we treat the $Y$ error as the correlated $Z$ operator on the two ancilla qubits.
In this way, the depolarizing noise is translated into 
correlated diagonal errors $\tilde{\mathcal{E}}_{1}$ on the ancilla qubits as follows:
\begin{align}
  \tilde{\mathcal{E}}_{1} = \pqty{1-\frac{3}{4}p}\bqty{I\otimes I} + \frac{p}{4}\pqty{\bqty{Z_1}  + \bqty{Z_2} + \bqty{Z_1Z_2}}.
\end{align}
This gives us the noisy resource state as
\begin{align}
  \tilde{\mathcal{E}}_{1}\pqty{\rho} = \pqty{1-\frac{3}{4}p}\rho+ \frac{p}{4}\pqty{\bqty{Z_1}  + \bqty{Z_2} + \bqty{Z_1Z_2}}\rho.
\end{align}

The noisy resource state obtained above is 
equivalent to the Choi state of the quantum channel $\mathcal{E}_\text{depo1} \circ \mathcal{U}$,
where $\mathcal{U}\pqty{\rho} \coloneqq U\rho U^\dagger$ up to  a local Clifford gate:
\begin{align}
  \tilde{\mathcal{E}}_{1}
  \pqty{\rho _{{\textrm{CZ}}\ket{U}\ket{+}}}
  =  [I] \otimes \left( [H] \circ \mathcal{E}_\text{depo1} \circ [U]\right) \pqty{ \rho _{\ket{\Psi ^{+}}}},
\end{align}
where
\begin{align}
\rho _{\pqty{\ket{\cdots}}} = \op{\cdots},\notag\\
\ket{\Psi ^{+}} \coloneqq \frac{\ket{00} + \ket{11}}{\sqrt{2}}, 
\end{align}
and CZ indicates the controlled-$Z$ gate.
More generally, for a diagonal single-qubit gate $U$ and single-qubit stochastic Pauli noise $\mathcal{E}$,
the classical simulation cost via the noise teleportation is quantified by a ROM of
\begin{align}
  [I] \otimes  \left(\mathcal{E} \circ [U] \right) \rho _{|\Psi ^{+}\rangle}.
\end{align}

Recently, a similar measure for a quantum channel was developed in Ref.~\cite{seddonQuantifyingMagicMultiqubit2019},
which is called the ``channel robustness."
The channel robustness is also calculated from the ROM of the Choi state except the stabilizer states used to decompose the Choi state must satisfy the trace-preserving condition.
Seddon and Campbell~\cite{seddonQuantifyingMagicMultiqubit2019} show that for any $n$-qubit CPTP (completely positive trace-preserving) maps $\mathcal{E}$, ROM of the Choi state $\mathcal{R}\pqty{\rho_{\mathcal{E}}}$, and channel robustness $\mathcal{R}_*\pqty{\mathcal{E}}$, the following inequality holds:
\begin{align}
  \mathcal{R}\pqty{\mathcal{\rho}_{\mathcal{E}}} \leq \mathcal{R}_*\pqty{{\mathcal{E}}},
\end{align}
where
\begin{align}
  \rho_\mathcal{E} &\coloneqq \pqty{\mathcal{E} \otimes I^{\otimes n}} \op{\Omega_n}, \\
  \ket{\Omega_n} &\coloneqq \frac{1}{\sqrt{2^n}}\sum_{j=0}^{2^n -1} \ket{j}\ket{j}.
\end{align}
We confirmed that the ROM of $\rho = \tilde{\mathcal{E}}_1\pqty{\op{T+}}$, which is the resource state of the $T$ gate followed by the depolarizing noise, is equal to its channel robustness by numerical calculations.
However, the ROM of the product state $\rho^{\otimes n}$ may be lower than the channel robustness of $\rho^{\otimes n}$.
This is because the stabilizer states that do not result in trace-preserving operations are used in the stabilizer decomposition when a resource state is multiple copies of a state.

\subsection{Noise fusion}
\label{subsec:synthesize_noise}
\begin{figure}[tb]
  \begin{subfigure}[t]{\linewidth}
    \centering
    \caption{}
    \label{subfig:not_synthesized_noise}
    \includegraphics[scale=1.5]{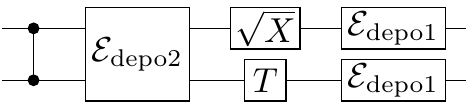}
  \end{subfigure}
  \begin{subfigure}[t]{\linewidth}
    \centering
    \caption{}
    \label{subfig:synthesized_noise}
    \includegraphics[scale=1.5]{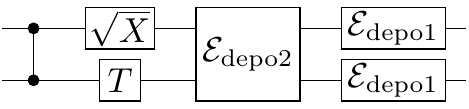}
  \end{subfigure}
  \caption{Noise fusion. \subref{subfig:not_synthesized_noise} Depolarizing noise occurs at every gate.
  \subref{subfig:synthesized_noise} Noise is collected around the $T$ gate to reduce the simulation cost of the $T$ gate.}\label{fig:schematic_noise_synthesization}
\end{figure}
In general, a quantum circuit consists of Clifford and non-Clifford gates.
Both of these gates are subject to noise.
Since Clifford gates can be simulated efficiently,
noise on the Clifford gates should not directly reduce the classical simulation costs for stabilizer-state sampling.
To reduce the ROM of a resource state, we construct a method to merge the noise on Clifford gates with the noise on a neighboring non-Clifford gate as follows.

Consider a quantum circuit consisting of arbitrary single-qubit and two-qubit gates, 
followed by single-qubit and two-qubit depolarizing noise,
respectively.
The single-qubit depolarizing noise and two-qubit depolarizing noise are given by
\begin{align}
  \mathcal{E}_\text{depo1}
  &\coloneqq
  \pqty{1-\frac{3}{4}p} \bqty{I} + \frac{p}{4}\sum  _{A \in \Bqty{X,Y,Z}} \bqty{A},
  \\
  \mathcal{E}_\text{depo2}
  &\coloneqq
  \pqty{1-\frac{15}{16}p} \bqty{I^{\otimes 2}} + \frac{p}{16} \sum_{\mathclap{\pqty{A, B}\neq\pqty{I, I}}} \; \bqty{A\otimes B}.
\end{align}
Let $U_1$ and $U_2$ be arbitrary single-qubit and two-qubit gates, respectively.
The unitary operator $U_1 \pqty{U_2}$ commutes with the depolarizing channel $\mathcal{E}_\text{depo1} \pqty{\mathcal{E}_\text{depo2}}$ as
\begin{align}
  \mathcal{E}_\text{depo1} \circ \bqty{U_1} &= \bqty{U_1} \circ \mathcal{E}_\text{depo1},
  \\
  \mathcal{E}_\text{depo2} \circ \bqty{U_2} &= \bqty{U_2} \circ \mathcal{E}_\text{depo2}. \label{eq:com_two_depo}
\end{align}
For example, consider the circuit shown in \cref{subfig:not_synthesized_noise}.
The two-qubit depolarizing noise forward is commuted with use of \cref{eq:com_two_depo}, which merges the single-qubit depolarizing noise and the two-qubit depolarizing noise [\cref{subfig:synthesized_noise}].
In general, for a given noise channel 
\begin{align}
  \mathcal{E} = \sum _i [E_i],
\end{align}
by replacing $\mathcal{E}$ with a unitary gate $U$,
we have
\begin{align}
  \mathcal{E} \circ [U]  =  [U]  \circ \mathcal{E}',
\end{align}
where 
\begin{align}
  \mathcal{E}' \coloneqq \sum _i \bqty{U^{\dag} E_{i} U}.
\end{align}
If channel $\mathcal{E}$ is a stochastic Pauli channel and if $U$ is a gate in the third or lower level of the Clifford hierarchy, including the $T$ gate, then
$\mathcal{E}'$ is a stochastic Clifford channel.
In this case, the merging process reduces the ROM of the resource state.
For clarity, we consider the simplest case, the depolarizing noise, below.

\subsection{Illustrative example of noise teleportation and noise fusion}
\begin{figure}[tb]
  \centering
  \begin{subfigure}[t]{\linewidth}
    \centering
    \caption{}
    \label{subfig:T_and_H_gate_with_noise}
    \includegraphics[scale=1.5]{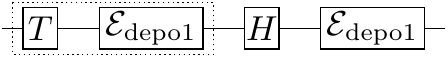}
  \end{subfigure}
  \begin{subfigure}[t]{\linewidth}
    \centering
    \caption{}
    \label{subfig:T_and_H_gate_with_noise_synthesizing}
    \includegraphics[scale=1.5]{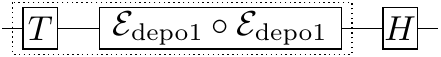}
  \end{subfigure}
  \caption{
    Circuits illustrating noise teleportation and noise fusion.
    \subref{subfig:T_and_H_gate_with_noise} $T$ gate and $H$ gate followed by a single-qubit depolarizing channel.
    \subref{subfig:T_and_H_gate_with_noise_synthesizing} Two single-qubit depolarizing channels are merged by the noise fusion.
  }
  \label{fig:illustrative_ex_of_noise_synthesization}
\end{figure}
\begin{figure}[tb]
  \centering
  \includegraphics[width=\linewidth]{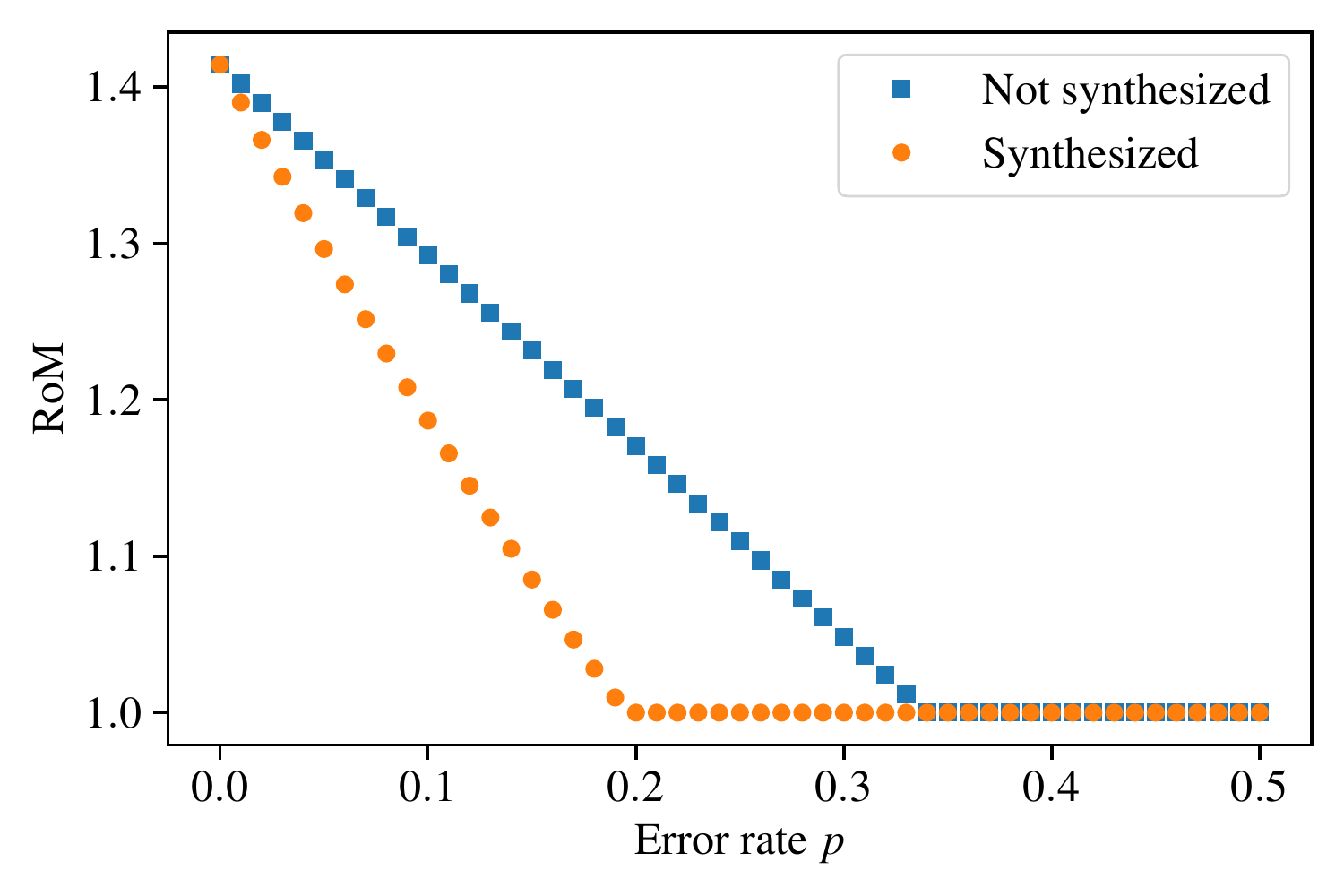}
  \caption{
    Comparison of the ROM with and without noise fusion.
    The horizontal axis shows the error rates of the single-qubit depolarizing noise.
    The vertical axis shows the ROM.
    The orange circles and blue squares correspond to the ROM defined in \cref{eq:rom_not_noise_fusion} with noise fusion and the ROM defined in \cref{eq:rom_noise_fusion} without noise fusion, respectively.
  }
  \label{fig:reduction_RoM_using_ns}
\end{figure}
To confirm that noise teleportation and noise fusion can reduce the ROM, 
we calculate the ROM of the noisy circuit shown in \cref{subfig:T_and_H_gate_with_noise}.
We assume that single-qubit depolarizing noise always occurs after a single-qubit gate.
The ROM of the resource state of the quantum channels surrounded by the dashed rectangle in \cref{subfig:T_and_H_gate_with_noise} is given as
\begin{align}
  \mathcal{R}\pqty{\tilde{\mathcal{E}_1}\pqty{\op{T+}}}\label{eq:rom_not_noise_fusion}.
\end{align}
We replace the single-qubit depolarizing noise backward using the noise fusion as shown in \cref{subfig:T_and_H_gate_with_noise_synthesizing}.
The ROM of the resource state of the quantum channels surrounded by the dashed rectangle in \cref{subfig:T_and_H_gate_with_noise_synthesizing} is given as
\begin{align}
  \mathcal{R}\pqty{\tilde{\mathcal{E}_1}\circ\tilde{\mathcal{E}_1} \pqty{\op{T+}}}. \label{eq:rom_noise_fusion}
\end{align}
The ROM of the resource state is calculated in both cases with noise fusion and without noise fusion.
The blue squares and orange circles in \cref{fig:reduction_RoM_using_ns} correspond to the ROM defined in \cref{eq:rom_not_noise_fusion,eq:rom_noise_fusion}, respectively.
The ROM shown in \cref{eq:rom_not_noise_fusion} is unity when the error rate $p$ is equal to or greater than 0.34.
On the other hand, the ROM shown in \cref{eq:rom_noise_fusion} is unity when the error rate $p$ is equal to or greater than 0.2.
Hence, noise fusion successfully reduces the ROM. 

\subsection{Reducing basis for stabilizer-state decomposition}
\label{subsec:reduce_stab_states}
\begin{figure}[tb]
  \centering
  \includegraphics[width=\linewidth]{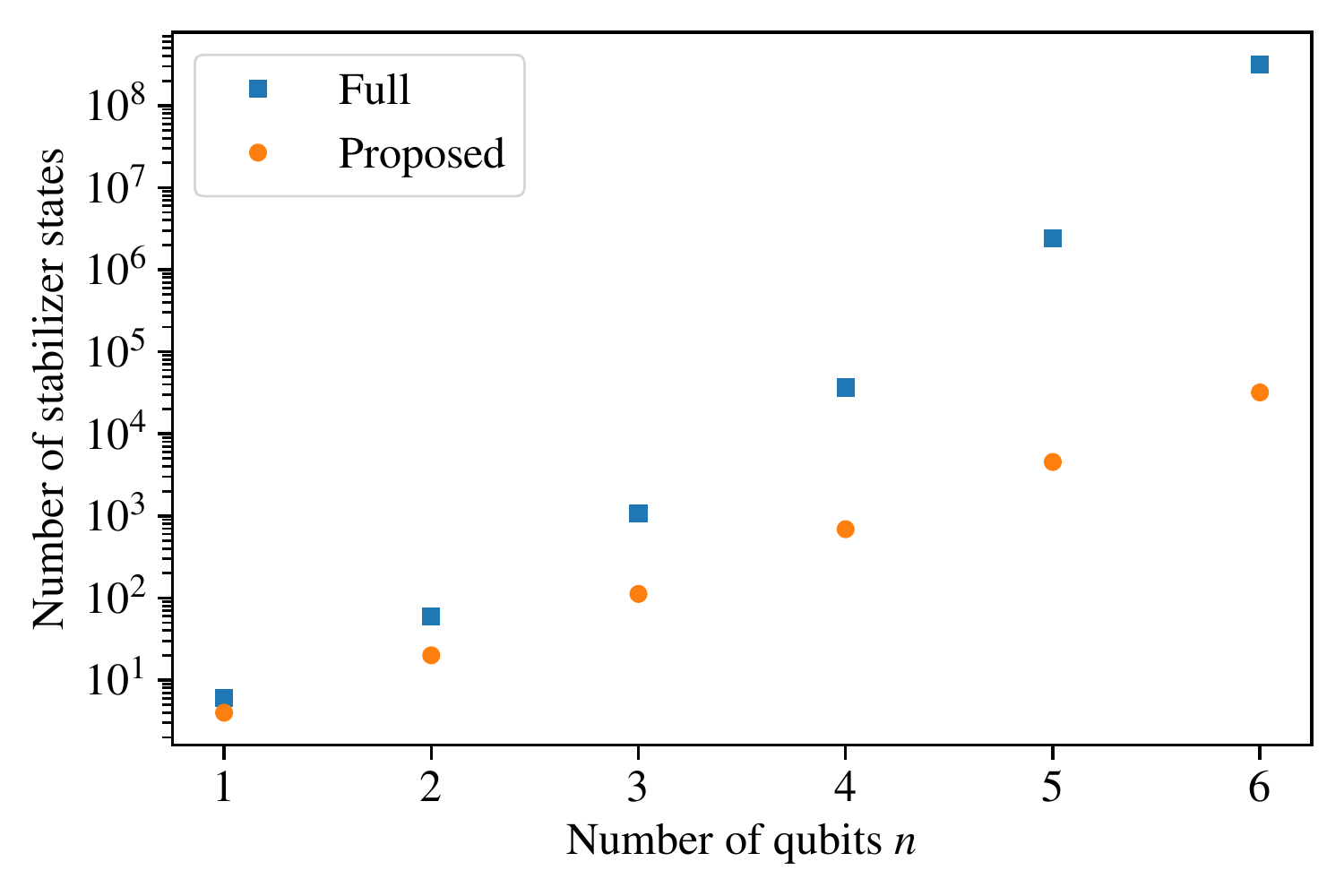}
  \caption{
    Comparison of the number of the stabilizer states used to calculate the exact ROM of $n$-fold copies of the magic state and the number of stabilizer states used to calculate its upper bound by our proposed method.
    The horizontal axis indicates the number of qubits.
    The vertical axis indicates the number of stabilizer states.
    Blue squares and orange circles correspond to the total number of stabilizer states and the number of stabilizer states needed to calculate the upper bound, respectively.
  }
  \label{fig:cp_num_stab_states}
\end{figure}
\begin{table}[tb]
  \centering
  \caption{
    Comparison of the exact ROM of $n$-fold copies of the magic state $\ket{T}$ and  their upper bounds calculated by our proposed method. 
  }
  \label{tab:comparison_RoM}
  \begin{tabular}{l|cc}
    \toprule
    $n$ &      Proposed method &     Exact \\
    \midrule
    1   &  1.414214 &  1.414214 \\
    2   &  1.747547 &  1.747547 \\
    3   &  2.218951 &  2.218951 \\
    4   &  2.862742 &  2.862742 \\
    5   &  3.689298 &  3.687052 \\
    \bottomrule
  \end{tabular}
\end{table}
Simulation costs can be reduced by decomposing $n$-fold copies of a resource state $\rho^{\otimes n}$ instead of decomposing $\rho$. 
Unfortunately,
the number of stabilizer states
used in the decomposition
grows superexponentially 
as the number of qubits increases ~\cite{aaronsonImprovedSimulationStabilizer2004,grossHudsonTheoremFinitedimensional2006}:
\begin{align}
  2^n\prod_{k=1}^n\pqty{2^k + 1}. \label{eq:number_of_stab}
\end{align}
Thus, it becomes intractable to calculate $\mathcal{R}\pqty{\op{T}^{\otimes n}}$ exactly as $n$ increases.
However, not all stabilizer states contribute to the decomposition of the 
tensor product of the magic states.
The decomposition of the magic state $\op{T}$ requires only
$X$ and $Y$ components of stabilizer states.
Hence, the stabilizer states whose stabilizer operators include $X$ and $Y$ operators 
are engaged in the decomposition.

Consider the upper bound of the ROM of two-copies of a magic state $\op{T}^{\otimes 2}$.
We use separable stabilizer states, all twofold tensor products of $\Bqty{\ket{\pm}, \ket{\pm i}}$ and the entangled stabilizer states stabilized by 
\begin{gather}
  \ev{XX, YY}, \quad \ev{-XX, -YY},\notag \\
  \ev{XY, YX}, \quad \ev{-XY, -YX}. \label{eq:entangled_stab_groups}
\end{gather}
Using these stabilizer states, we numerically find that the upper bound of the ROM of $\op{T}^{\otimes 2}$ is almost the same as the exact value.
On the basis,
we use the separable stabilizer states 
and bipartite entangled stabilizer states with respect to $X$ and $Y$ operators.
To decompose $\op{T}^{\otimes n}$ over certain stabilizer states, we use the stabilizer states stabilized by one of the stabilizer groups shown in \cref{eq:entangled_stab_groups}, for up to $\floor*{\frac{n}{2}}$ pairs.
Thus, the number of basis elements to decompose $\op{T}^{\otimes n}$ from \cref{eq:number_of_stab} is reduced to
\begin{align}
  4^n n! \sum_{k=0}^{\floor*{n/2}} \Bqty{8^k k! \pqty{n-2k}!}^{-1}.
\end{align}
\Cref{fig:cp_num_stab_states} compares the number of the stabilizer states used to calculate the ROM of $n$-fold copies of the magic state $\mathcal{R}\pqty{\op{T}^{\otimes n}}$ and its upper bound by our proposed method.
The horizontal axis indicates the number of stabilizer states, while the vertical axis indicates the number of qubits.
The blue squares and orange circles correspond to the total number of stabilizer states and the number of stabilizer states used by our proposed method, respectively.
Our method uses fewer stabilizer states than the total number of stabilizer states.
\Cref{tab:comparison_RoM} compares the upper bounds of the ROM of $n$-fold copies of the magic state for our proposed method with the exact ROM calculated with all stabilizer states.
The upper bound of the ROM is sufficiently close to the exact ROM up to four qubits.
For $n=5$, the upper bound of the ROM is slightly larger than the exact ROM.
Therefore, our method provides a reasonable upper bound of the ROM.

This method is used in Sect.~\ref{sec:comparison_two_methods} in the caluclation of the ROM of twofold copies of the resource state of the depolarized $T$ gates.
A more-efficient and more-precise method for calculating the upper bound of the ROM of $n$-fold copies of the magic state was proposed in Ref.~\cite{heinrichRobustnessMagicSymmetries2019},
but the noisy case was not considered there.

\section{Comparison of sampling-based simulation algorithms}\label{sec:comparison_two_methods}
\begin{figure}[tb]
  \begin{subfigure}[t]{\linewidth}
    \centering
    \caption{}
    \label{subfig:unit1}
    \includegraphics[scale=1.5]{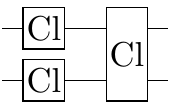}
  \end{subfigure}
  \begin{subfigure}[t]{\linewidth}
    \centering
    \caption{}
    \label{subfig:unit2}
    \includegraphics[scale=1.5]{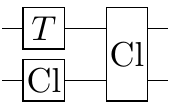}
  \end{subfigure}
  \begin{subfigure}[t]{\linewidth}
    \centering
    \caption{}
    \label{subfig:unit3}
    \includegraphics[scale=1.5]{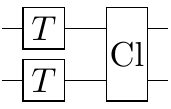}
  \end{subfigure}
  \caption{
    Three patterns of the unit cell constituting our RQCs:
    \subref{subfig:unit1} unit cell 1,
    \subref{subfig:unit2} unit cell 2, and
    \subref{subfig:unit3} unit cell 3.
    Cl denotes a Clifford gate.
  }
  \label{fig:unit_cell_circs}
\end{figure}
Here we compare the two sampling-based algorithms
by calculating the simulation cost concretely for RQCs,
which are similar to the circuits used in the quantum computational supremacy experiment~\cite{aruteQuantumSupremacyUsing2019},
consisting of Clifford and $T$ gates followed by 
single-qubit or two-qubit depolarizing noise.
We construct RQCs by alternately applying single-qubit gates chosen from $\Bqty{\sqrt{X}, \sqrt{Y}, T}$ to each qubit, and then applying two-qubit Clifford gates to nonoverlapping nearest-neighbor pairs of two qubits.
We call the sequential operations of applying single-qubit and two-qubit gates in turn a ``cycle".
After a certain number of cycles, all qubits are measured to obtain an expectation value of Pauli operator $P$.

The expectation value of $P$ with respect to an output state of a RQC becomes exponentially close to zero as the depth of a RQC increases. However, this is not the point in our comparison.
The RQCs are just chosen as a representative of hardware-efficient quantum circuits that are expected to be used in near-term quantum devices.
Regarding a single-qubit gate, the $T$ gate is the most adversarial in classical simulation, since the ROM of the associated resource state is the highest. 
We expect that the ROM of the hardware-efficient ansatz with the same arrangement of qubits will be much smaller. Therefore, our result will also give some insights into classical simulatability of such an ansatz.

The unit cell of these RQCs comprises two single-qubit gates followed by one two-qubit Clifford gate.
We assume that single-qubit and two-qubit depolarizing noise occurs after each single-qubit gate and two-qubit gate, respectively.
\Cref{fig:unit_cell_circs} shows the three patterns of the unit cell.
Unit cell 1 consists of only Clifford gates.
Unit cell 2 consists of one $T$ gate and two Clifford gates, while unit cell 3 consists of two $T$ gates and one Clifford gate.
This structure can provide the upper bound for the simulation cost of a noisy RQC.

In the case of Heisenberg propagation, the upper bound of the simulation cost of n RQC can be estimated from the simulation costs of unit cells 1, 2, and 3 straightforwardly by \cref{eq:sim_costs_Pauli}.
First we calculate the channel stabilizer norm using the Pauli transfer matrix (PTM) of a quantum channel~\cite{rallSimulationQubitQuantum2019}.
For a quantum channel $\Lambda$ that takes $n$ qubits to $n$ qubits, the corresponding PTM is defined as
\begin{align}
  \pqty{R_\Lambda}_{ij} \coloneqq \frac{1}{2^n}\Tr\bqty{P_i \Lambda\pqty{P_j}}.
\end{align}
Combining the definition of the PTM with that of the channel stabilizer norm [see \cref{eq:ch_st_norm}], we obtain
\begin{align}
  \mathcal{D}\pqty{\Lambda^\dag} = \norm{R_\Lambda}_\infty,
\end{align}
where $\norm{\cdot}_\infty$ is the largest row's L1 norm of the matrix $R_\Lambda$.
The PTMs of the $T$ gate, single-qubit depolarizing noise, and two-qubit depolarizing noise are given by
\begin{align}
  R_T &= \bmqty{
    1 & 0 & 0 & 0\\
    0 & \frac{1}{\sqrt{2}} & -\frac{1}{\sqrt{2}} & 0\\
    0 & \frac{1}{\sqrt{2}} & \frac{1}{\sqrt{2}} & 0\\
    0 & 0 & 0 & 1
  },\label{eq:PTM_T}\\
    R_\text{depo1} &= \text{diag}\pqty{1, 1-p_1, 1-p_1, 1-p_1},\label{eq:PTM_depo1} \\
    R_\text{depo2} &= \text{diag}\pqty{1, 1-p_2,\ldots,1-p_2},\label{eq:PTM_depo2}
  \end{align}
respectively.
The PTMs of Clifford gates do not need to be considered because they are signed permutation matrices and the depolarizing noise is symmetric.
Because unit cell 1 includes only Clifford gates, its channel stabilizer norm is unity.
By multiplying the PTMs in \cref{eq:PTM_T,eq:PTM_depo1,eq:PTM_depo2},
the PTMs of unit cells 2 and 3 are written, respectively, as
\begin{align}
  R_\text{unit 2} &= \pqty{R_T \otimes I_{4}} \pqty{R_\text{depo1} \otimes I_{4}} \pqty{I_4 \otimes R_\text{depo1}} R_\text{depo2}, \notag\\
  R_\text{unit 3} &= \pqty{R_T \otimes I_{4}} \pqty{I_4 \otimes R_T}\notag \\
                 &\quad \pqty{R_\text{depo1} \otimes I_{4}} \pqty{I_4 \otimes R_\text{depo1}} R_\text{depo2}. \notag
\end{align}
The channel stabilizer norms of unit cells 2 and 3 are given, respectively, by
\begin{align}
  \mathcal{D}_{\text{unit 2}} &= \max\pqty{1, \sqrt{2}\pqty{p_1 - 1}\pqty{p_2 -1}},\notag\\
  \mathcal{D}_{\text{unit 3}} &= \max\pqty{1, \sqrt{2}\pqty{p_1 - 1}\pqty{p_2 - 1}, -2\pqty{p_1 - 1}^2 \pqty{p_{2} - 1}}\notag.
\end{align}

On the other hand, 
for stabilizer-state sampling, 
we calculate the ROM of unit cells 2 and 3 with the depolarizing noise.
The error-free resource states of unit cells 2 and 3 are given by 
\begin{align}
\rho_{\ket{T+++}} = \op{T+++},\notag\\
\rho_{\ket{T+T+}} = \op{T+T+},
\end{align}
respectively.
We add the correlated diagonal noise on them 
to consider the depolarizing noise, and calculate the ROM of depolarized unit cells 2 and 3.
The noisy resource states of unit cells 2 and 3 are given, respectively, by
\begin{align*}
  \rho_\text{unit2} &= \tilde{\mathcal{E}}_2 \circ \pqty{\tilde{\mathcal{E}}_1 \otimes I} \circ \pqty{I \otimes \tilde{\mathcal{E}}_1} \pqty{\rho_{\ket{T+++}}}, \\
  \rho_\text{unit3} &= \tilde{\mathcal{E}}_2 \circ \pqty{\tilde{\mathcal{E}}_1 \otimes I} \circ \pqty{I \otimes \tilde{\mathcal{E}}_1} \pqty{\rho_{\ket{T+T+}}},
\end{align*}
where
\begin{align}
  \tilde{\mathcal{E}}_2 &\coloneqq \pqty{1-\frac{15}{16}p}\bqty{I^{\otimes 4}} + \frac{p}{16}\sum_{(A,B) \in S} A \otimes B, \label{eqref:noisy_unit_resoruce} \\
  S &\coloneqq \Bqty{\pqty{A,B}|A,B\in \Bqty{I^{\otimes 2}, Z_1, Z_2, Z_1 Z_2}} \setminus \Bqty{\pqty{I^{\otimes 2}, I^{\otimes 2}}}\notag
\end{align}
To reduce the simulation cost, we also calculate the upper bound for the ROM of twofold copies of $\rho_{\text{unit$i$}}$ ($i=2,3$).
We call stabilizer-state sampling with twofold copies of a resource state ``optimized stabilizer-state sampling."
In this case, the RoM per unit cell $i$ is given by  $\Bqty{\mathcal{R}\pqty{\rho_\text{unit$i$}^{\otimes 2}}}^{\frac{1}{2}}$.
The corresponding simulation cost of unit $i$ is proportional to $\mathcal{R}\pqty{\rho_\text{unit$i$}^{\otimes 2}}$.
Throughout the optimization, we use the basis reduction method as explained in Section~\ref{subsec:reduce_stab_states} because the decomposition of an eight-qubit resource state is intractable for a current classical computer.
To decompose the resource states of depolarized Clifford+$T$ circuits over certain stabilizer states, we use the stabilizer states stabilized by one of the stabilizer groups shown in \cref{eq:entangled_stab_groups} for the qubit pairs applied by the $T$ gate in the resource state.
For the other qubits, which are not applied by the $T$ gate in the resource state, we use the stabilizer state $\ket{+}$ or $\ket{-}$.

The ratio of the simulation cost of a circuit for calculating the expectation value of $P$ via Heisenberg propagation to that via stabilizer-state sampling is given by
\begin{align*}
  \frac{N_\text{Heisenberg}}{N_\text{stabilizer}} = \pqty{\frac{\prod_{i=1}^d \mathcal{D} \pqty{\Lambda_i^\dag}}{\mathcal{R}\pqty{\rho}}}^2,
\end{align*}
where we use $\mathcal{D}\pqty{P} = 1$.
Therefore, we can compare the simulation costs for the two sampling-based classical simulation methods by calculating the ROM and the product of channel stabilizer norms.

\begin{figure}[tb]
  \centering
  \begin{subfigure}[t]{\linewidth}
    \caption{}
    \label{subfig:unit2_cost}
    \includegraphics[keepaspectratio, width=\linewidth]{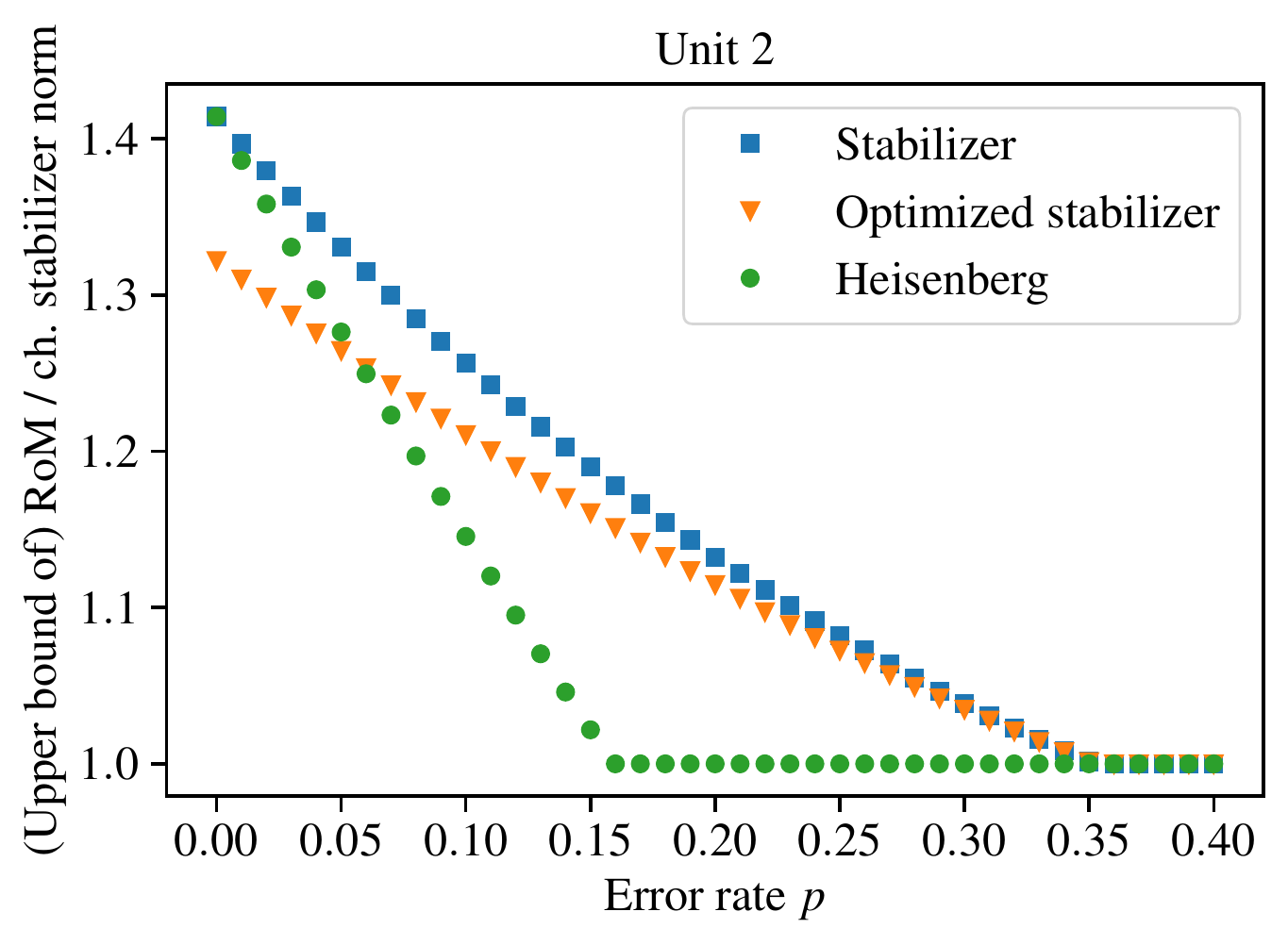}
  \end{subfigure}
  \begin{subfigure}[t]{\linewidth}
    \centering
    \caption{}
    \label{subfig:unit3_cost}
    \includegraphics[keepaspectratio, width=\linewidth]{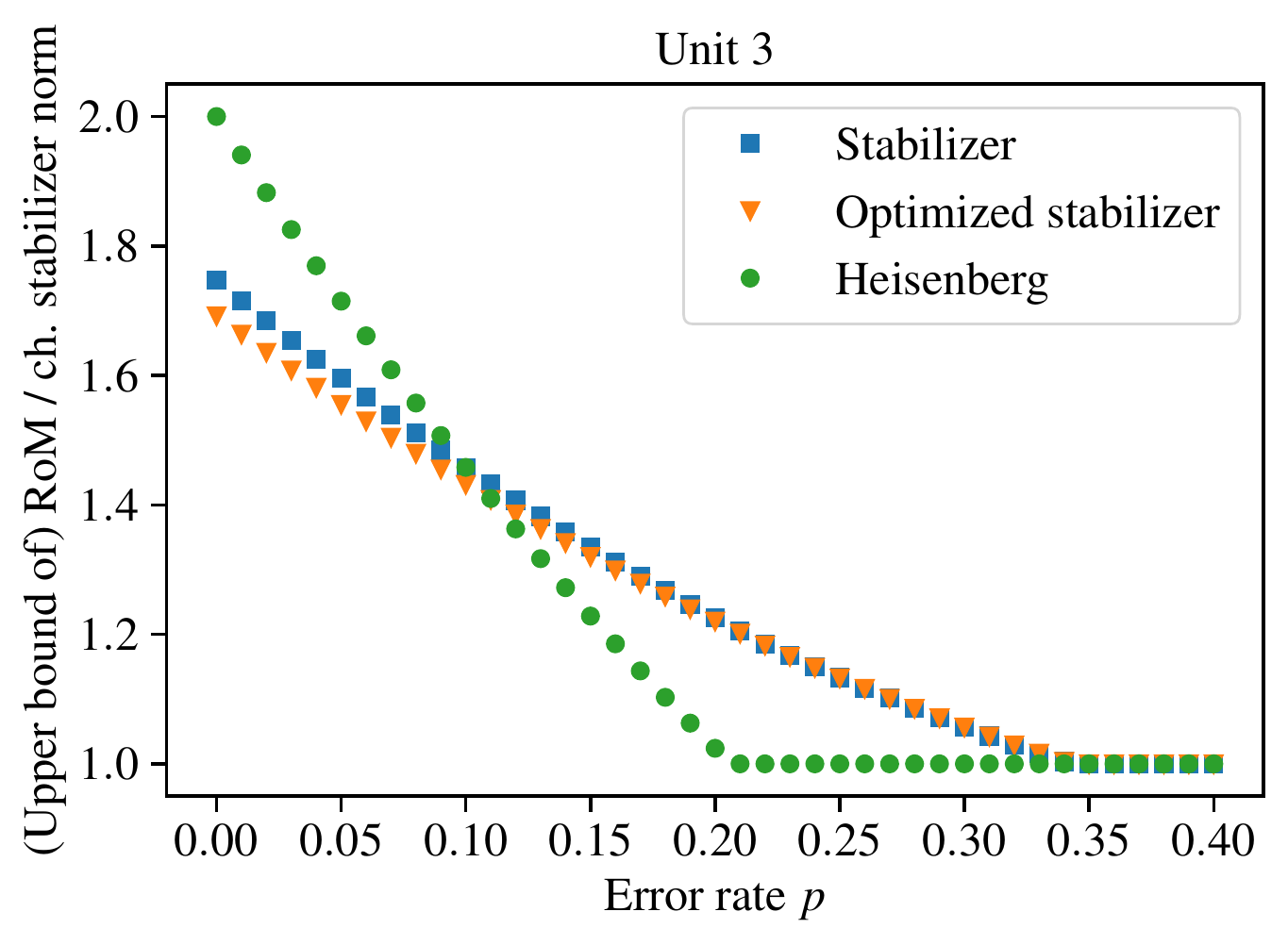}
  \end{subfigure}
  \caption{Comparison of the simulation costs of faulty unit cell 2 \subref{subfig:unit2_cost} and faulty unit cell 3 \subref{subfig:unit3_cost}.
    The horizontal axis displays the error rates of the single-qubit and two-qubit depolarizing noise.
    The vertical axis displays the ROM, the upper bound of the ROM, or the channel (ch.) stabilizer norms, each of which quantifies the simulation cost of the corresponding method under consideration.
    The Blue squares, orange triangles, and green circles stand for simulation costs for stabilizer-state sampling $\bqty{\mathcal{R}\pqty{\rho_\text{unit$i$}}}$, optimized stabilizer-state sampling $\pqty{\Bqty{\mathcal{R}\pqty{\rho_\text{unit$i$}}^{\otimes 2}}^{\frac{1}{2}}}$, and Heisenberg propagation $\pqty{\mathcal{D}_\text{unit$i$}}$, respectively, where $i=2,3$.
  }
  \label{fig:unit_cost}
\end{figure}
\Cref{fig:unit_cost} (a) and (b) show the resultant simulation costs of unit cells 2 and 3, respectively.
For simplicity, we assume that $p_1 = p_2 = p$.
The blue squares, orange triangles, and green circles stand for
simulation costs for 
stabilizer-state sampling $\bqty{\mathcal{R}\pqty{\rho_\text{unit$i$}}}$, 
optimized stabilizer-state sampling $\pqty{\Bqty{\mathcal{R}\pqty{\rho_\text{unit$i$}}^{\otimes 2}}^{\frac{1}{2}}}$, and 
Heisenberg propagation $\pqty{\mathcal{D}_{\text{unit$i$}}}$, respectively, where $i=2,3$.
In the low-noise region, 
the simulation costs for stabilizer-state sampling are much smaller than those of Heisenberg propagation.
This is because the simulation costs are reduced by optimization of the stabilizer-state decomposition by use of multiple copies of a resource state.
However, this is not the case for Heisenberg propagation.
Suppose a channel stabilizer norm is calculated for a merged channel over several patterns of the unit cell.
In this case, the simulation cost should be further reduced, whereas the dimensions of the PTM increase exponentially.
However, we do not consider merging multiple patterns of the unit cell because the simulation cost of Heisenberg propagation depends explicitly on quantum channels.
The dependence on quantum channels makes optimization of the channel stabilizer norm by merging the quantum channels problematic due to the enormous number of possible merging patterns.

In the high-noise region, Heisenberg propagation provides lower simulation costs.
This is attributed to the difference in the effect of the depolarizing noise in both algorithms.
To explain this in detail, we first consider the simulation cost of $\mathcal{E}_A \circ \bqty{T}$ for stabilizer-state sampling, where we set $\mathcal{E}_A$ $(A=X,Y,Z)$ a single-qubit general dephasing noise:
\begin{align}
    \mathcal{E}_\text{$A$} \coloneqq \pqty{1-p}\bqty{I} + p\bqty{A}.
\end{align}
We numerically calculate the ROM of the resource states of $\mathcal{E}_\text{$A$} \circ \bqty{T}$ for $A=X$, $Y$, and $Z$.
The ROM of the resource state of $\mathcal{E}_\text{$Z$} \circ \bqty{T}$ is smallest for all $p$.
As a result, for stabilizer-state sampling, the $Z$ error in the single-qubit depolarizing noise  has a larger impact on the simulation costs of noisy quantum circuits compared with other Pauli errors: the $X$ and $Y$ errors.
On the other hand, for Heisenberg propagation, the two Pauli errors in the depolarizing noise always help reduce the simulation costs.
Let $\Lambda_\text{Pauli}$ be the single-qubit stochastic Pauli noise
defined as follows:
\begin{align*}
  \Lambda_\text{Pauli} \coloneqq \pqty{1-\frac{p_X+p_Y+p_Z}{4}} \bqty{I} + \sum_{\mathclap{P=X,Y,Z}} \; \frac{p_P}{4} \bqty{P}.
\end{align*}
The PTM of $\Lambda_\text{Pauli} \circ \bqty{T}$ is given by
\begin{align*}
  \bmqty{
    1 & 0 & 0 & 0 \\
    0 & \frac{1}{\sqrt{2}}\pqty{1 - \frac{p_Y + p_Z}{2}} & \frac{1}{\sqrt{2}}\pqty{1 - \frac{p_Y + p_Z}{2}} & 0 \\
    0 & \frac{-1}{\sqrt{2}}\pqty{1 - \frac{p_X + p_Z}{2}} & \frac{1}{\sqrt{2}}\pqty{1 - \frac{p_X + p_Z}{2}} & 0 \\
    0 & 0 & 0 & 1-\frac{p_X + p_Y}{2}
  },
\end{align*}
so 
\begin{align}
    \mathcal{D}\pqty{\Lambda_\text{Pauli} \circ \bqty{T}} &= \max \bigg\{1, \sqrt{2}\pqty{1-\frac{p_X + p_Z}{2}},\notag\\
    &\quad \sqrt{2}\pqty{1-\frac{p_Y + p_Z}{2}}\bigg\}.
\end{align}
We see that the two Pauli errors, $\Bqty{X,Y}$ or $\Bqty{Y,Z}$, decrease the simulation cost of $\Lambda_\text{Pauli} \circ \bqty{T}$ for Heisenberg propagation.
This explains why the simulation costs for Heisenberg propagation decrease faster than for stabilizer-state sampling according to the error rate of the depolarizing noise.

In short, in \cref{fig:unit_cost}, there is a crossover between which algorithm is better.
In the low-noise region, stabilizer-state sampling outperforms Heisenberg propagation.
However, the opposite is observed in the high-noise region.
The crossover appears around $p \geq 0.05$ for unit cell 2 and $p \geq 0.11$ for unit cell 3.

\begin{figure*}[t]
  \begin{minipage}[b]{.49\textwidth}
    \begin{subfigure}[t]{.9\textwidth}
      \caption{}
      \label{subfig:sim_costs_random_circ_0}
      \includegraphics[keepaspectratio, width=\textwidth]{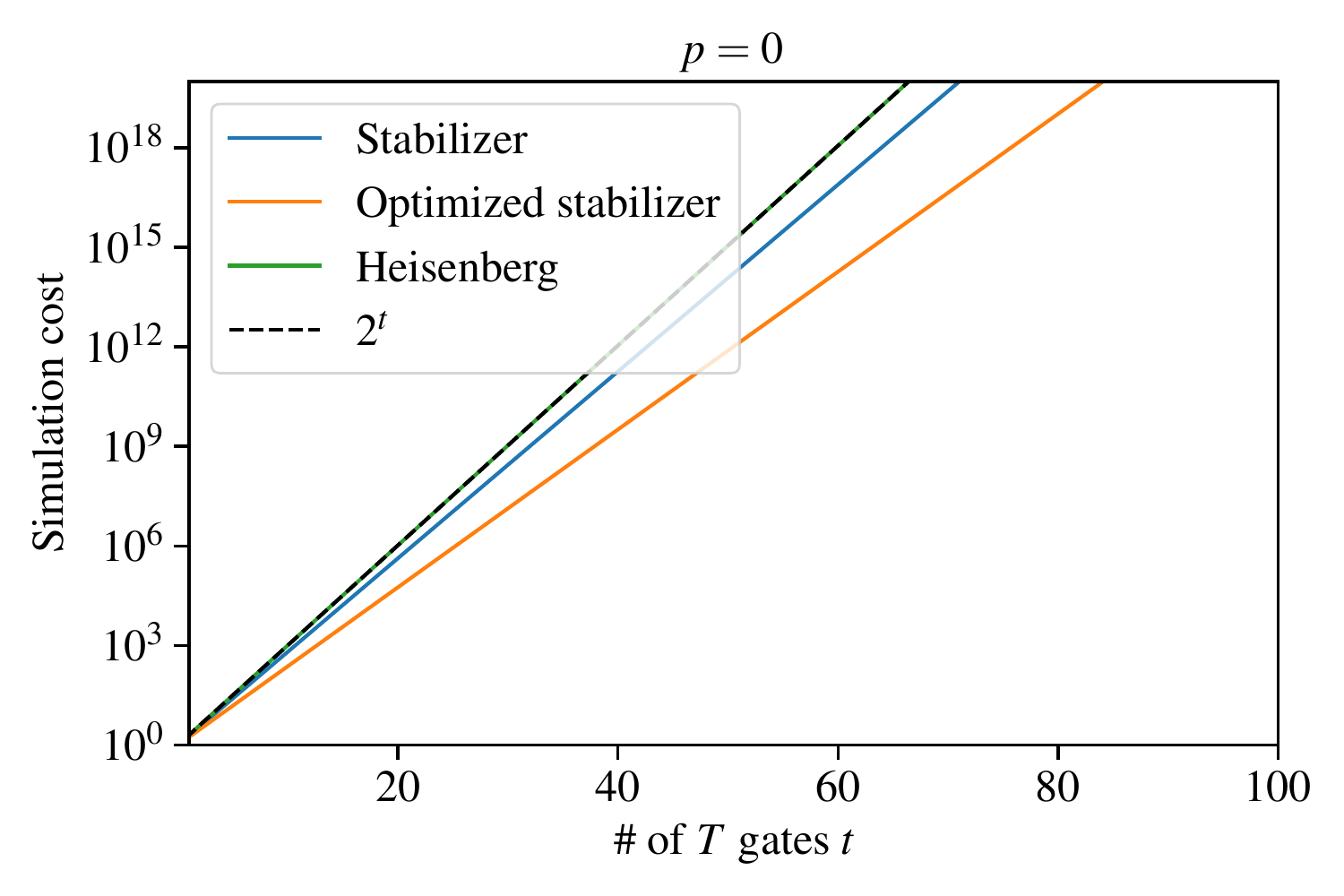}
    \end{subfigure}
  \end{minipage}
  \begin{minipage}[b]{.49\textwidth}
    \begin{subfigure}[t]{.9\textwidth}
      \caption{}
      \label{subfig:sim_costs_random_circ_005}
      \includegraphics[keepaspectratio, width=\textwidth]{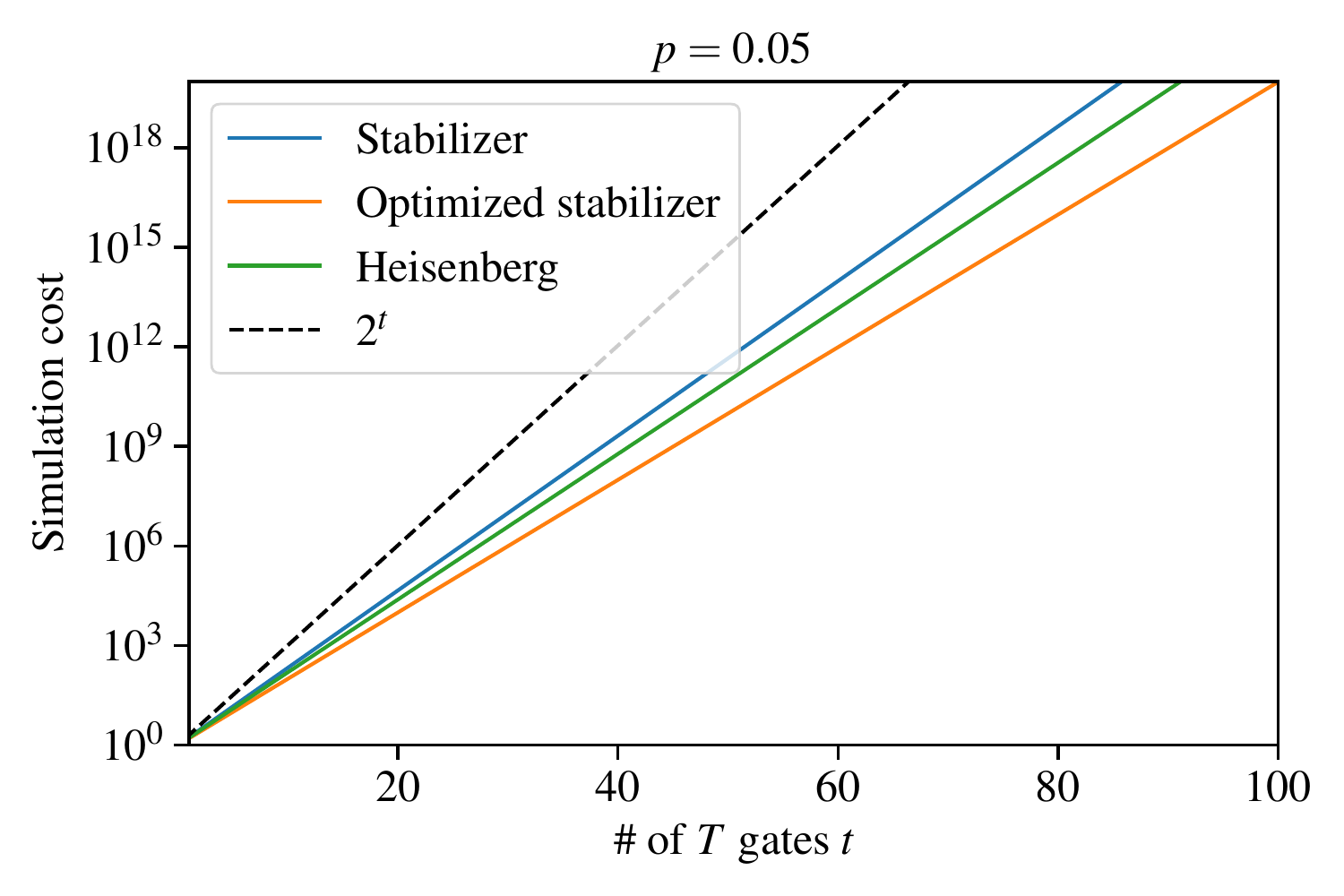}
    \end{subfigure}
  \end{minipage}\\
  \begin{minipage}[b]{.49\textwidth}
    \begin{subfigure}[t]{.9\textwidth}
      \caption{}
      \label{subfig:sim_costs_random_circ_01}
      \includegraphics[keepaspectratio, width=\textwidth]{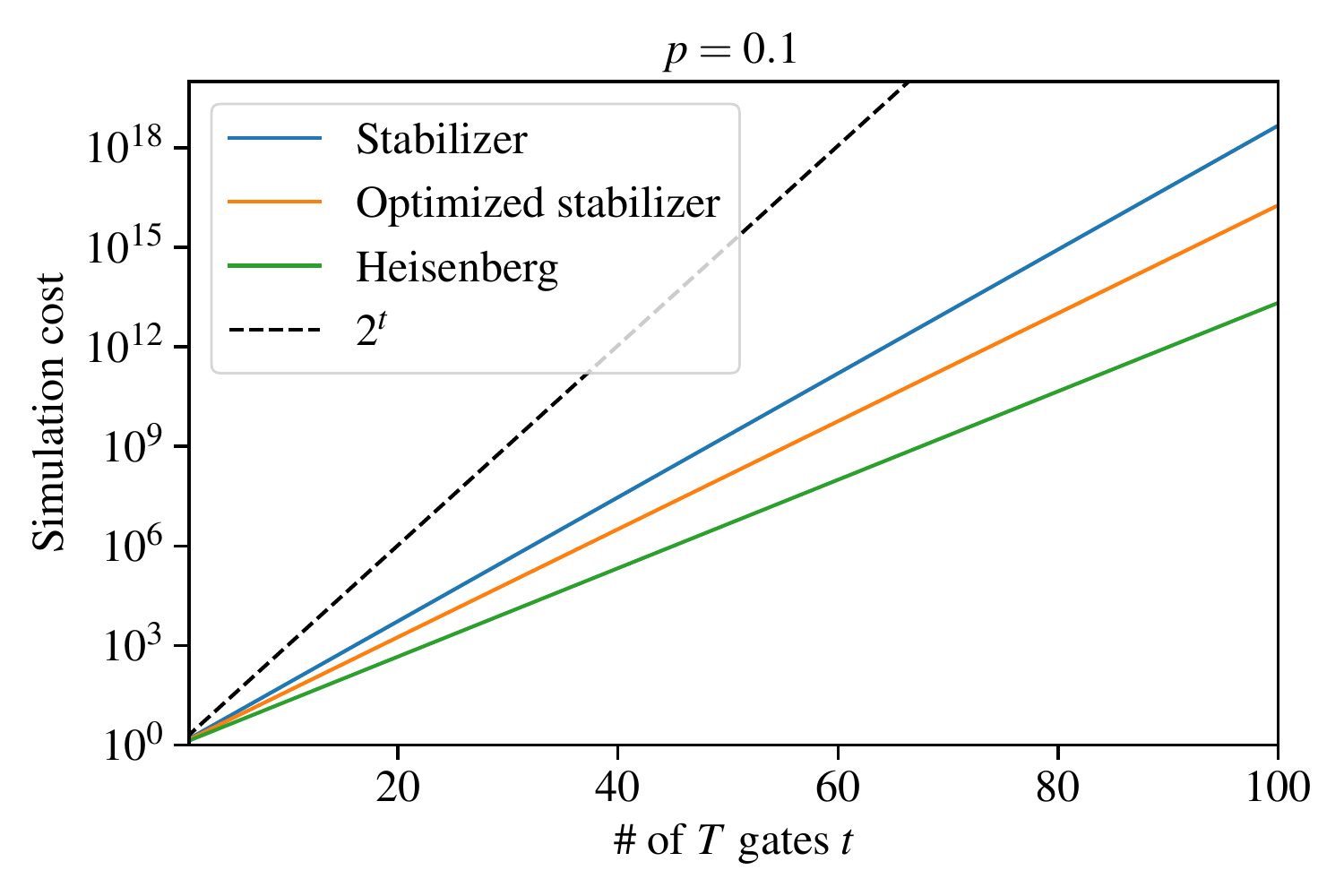}
    \end{subfigure}
  \end{minipage}
  \begin{minipage}[b]{.49\textwidth}
    \begin{subfigure}[t]{.9\textwidth}
      \caption{}
      \label{subfig:sim_costs_random_circ_015}
      \includegraphics[keepaspectratio, width=\textwidth]{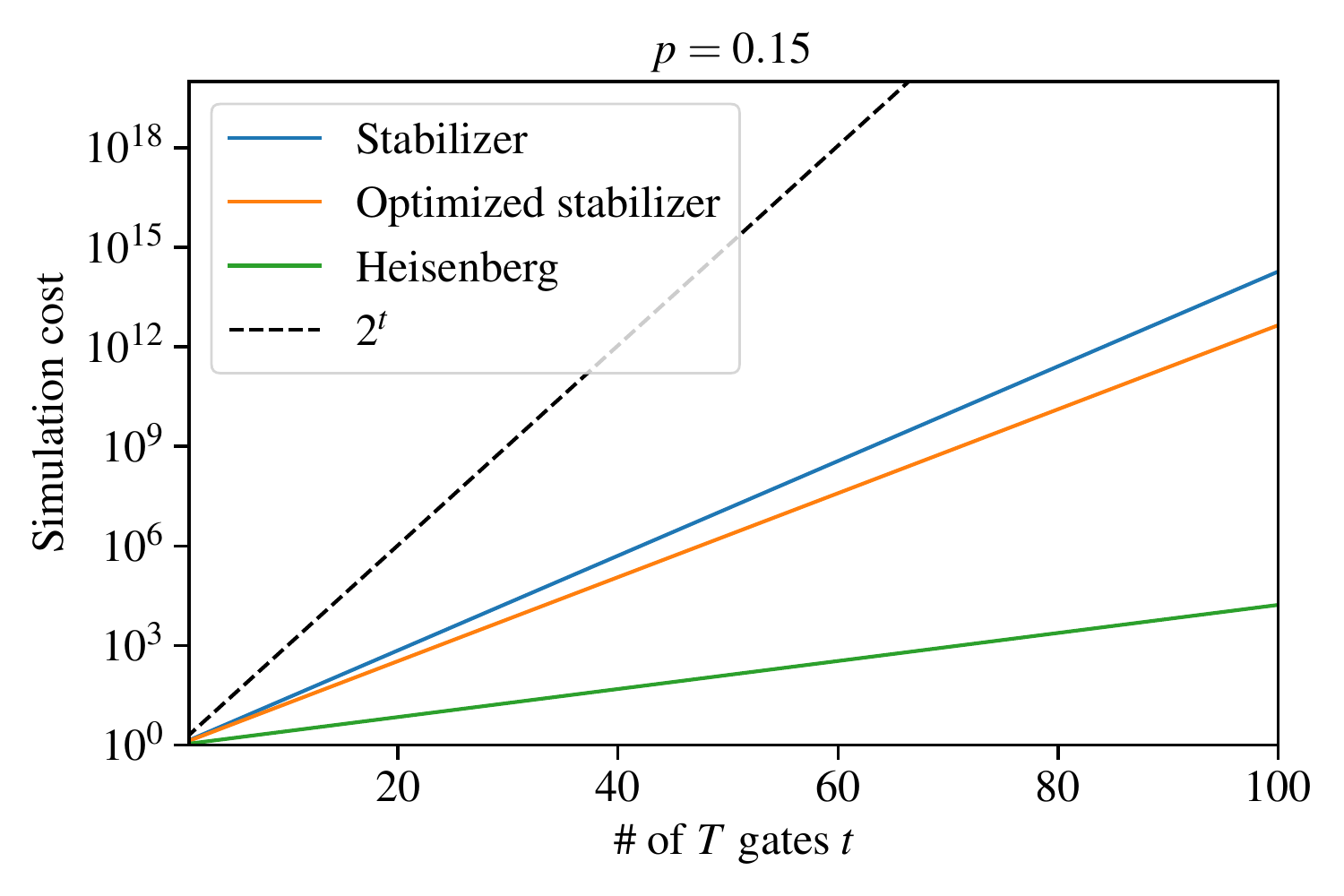}
    \end{subfigure}
  \end{minipage}

  \raggedright
  \begin{minipage}[b]{.49\textwidth}
    \begin{subfigure}[t]{.9\linewidth}
      \caption{}
      \label{subfig:sim_costs_random_circ_02}
      \includegraphics[keepaspectratio, width=\columnwidth]{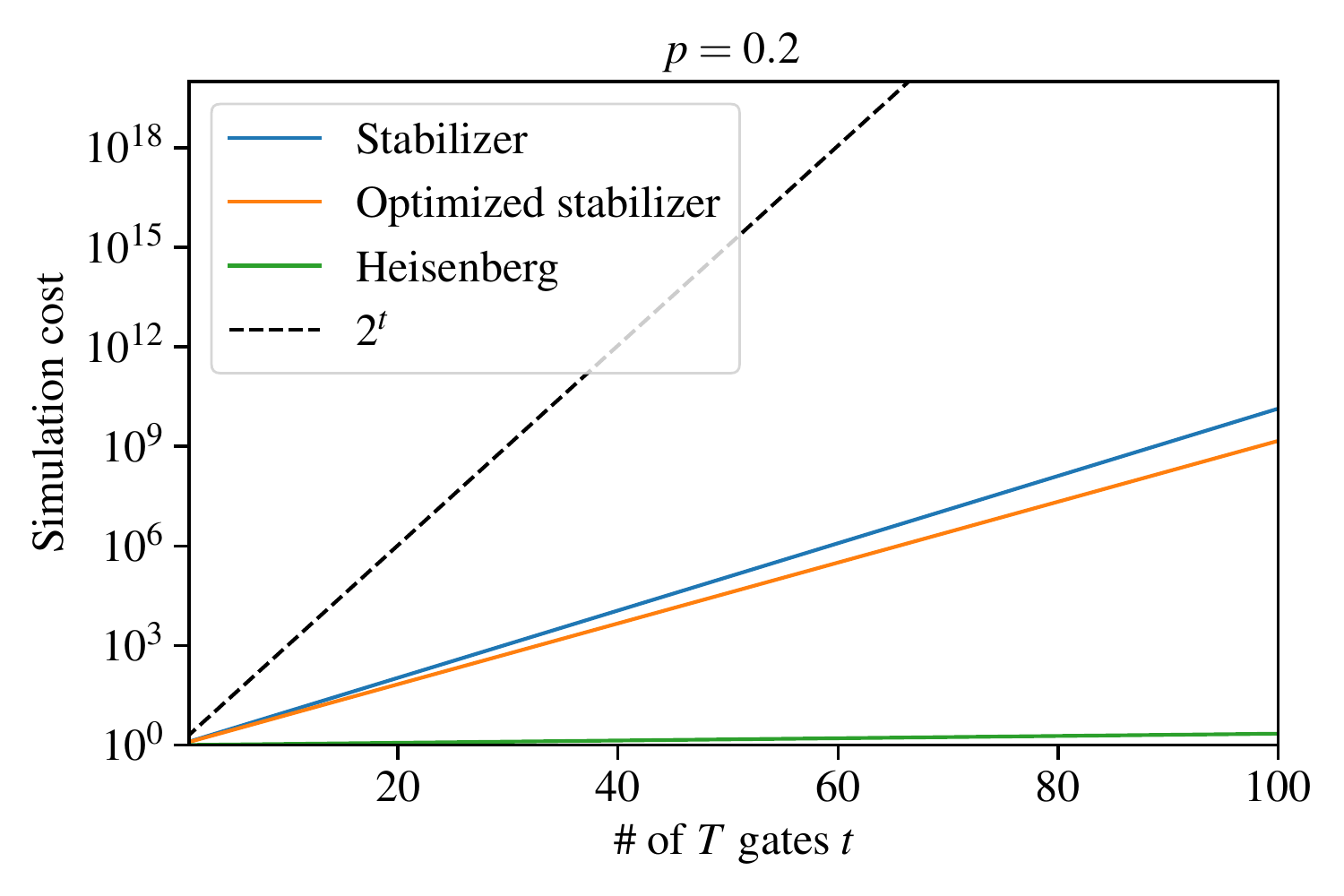}
    \end{subfigure}
  \end{minipage}
  \caption{
    Comparison of the simulation costs of depolarized RQCs with various error rates of the single-qubit and two-qubit depolarizing noise $p$ through stabilizer-state sampling and Heisenberg propagation.
    The horizontal axis shows the number of $T$ gates $t$ in the circuits.
    The vertical axis shows the squared products of the upper bounds of the ROM or those of the channel stabilizer norms.
    Both are proportional to the simulation costs.
    Blue, orange, and green lines correspond to the simulation costs as a function of $t$ for stabilizer-state sampling, optimized stabilizer-state sampling, and Heisenberg propagation, respectively.
    The simulation costs with $p=0, 0.05, 0.1, 0.15, 0.2$ are shown in 
    from \subref{subfig:sim_costs_random_circ_0} to \subref{subfig:sim_costs_random_circ_02}, respectively.
  }
  \label{fig:sim_costs_random_circ}
\end{figure*}
Next we calculate the total amount of simulation costs of the noisy RQCs on a rectangular lattice consisting of $m\times n$ qubits and $d$ cycles.
The total number of unit cells is roughly $D=mnd/2$.
The numbers of unit cells 1, 2, and 3 are roughly $\frac{4}{9}D$, $\frac{4}{9}D$, and $\frac{1}{9}D$, respectively,
since single-qubit gates are chosen randomly from $\Bqty{\sqrt{X},\sqrt{Y}, T}$.
The simulation costs for Heisenberg propagation are proportional to
\begin{align}
&\bigg[\Bqty{\max\pqty{1, \sqrt{2}\pqty{p-1}^2}}^{\frac{4D}{9}} \notag\\
&\times \Bqty{\max\pqty{1, \pqty{p-1}^2, -2\pqty{p-1}^3}}^{\frac{D}{9}}\bigg]^2.\notag
\end{align}
On the other hand, the simulation costs for stabilizer-state sampling are proportional to
\begin{align*}
  \bqty{\mathcal{R}\pqty{\rho_\text{unit2}}^\frac{4D}{9}\mathcal{R}\pqty{\rho_\text{unit3}}^\frac{D}{9}}^2.
\end{align*}
Since the number of $T$ gates $t$ is relevant in the comparison of the simulation costs
in both algorithms, 
we calculate the simulation costs as a function of $t$, which 
is related to $D$ by $\frac{3}{2}D$.
\Cref{fig:sim_costs_random_circ} shows the simulation costs of the two algorithms
for $p=0, 0.05, 0.1, 0.15, 0.2$.
The blue, orange, and green lines correspond to the simulation costs as a function of  $t$ for stabilizer-state sampling, optimized stabilizer-state sampling, and Heisenberg propagation, respectively.
If the error rate $p$ is low, the optimized stabilizer-state sampling is the best choice [\cref{subfig:sim_costs_random_circ_0,subfig:sim_costs_random_circ_005}], otherwise, Heisenberg propagation is superior to stabilizer-state sampling [\crefrange{subfig:sim_costs_random_circ_01}{subfig:sim_costs_random_circ_02}].

\begin{figure}[tb]
  \centering
  \includegraphics[keepaspectratio, width=\columnwidth]{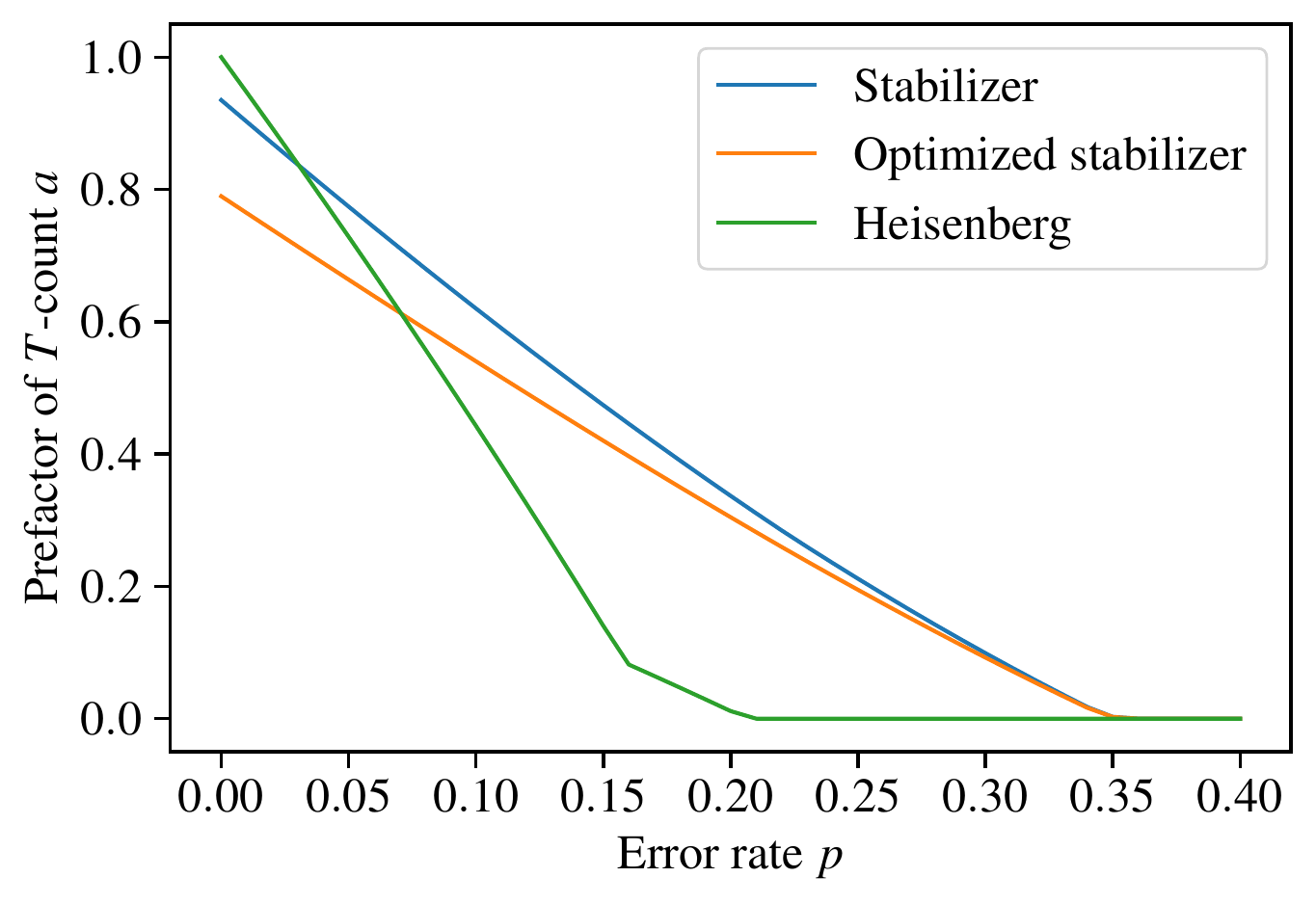}
  \caption{
    Scaling factors of the simulation costs as a function of the error rate $p$ of single-qubit and two-qubit depolarizing noise.
    The horizontal axis shows the error rates of the depolarizing noise $p$.
    The vertical axis shows the prefactor of the exponent $\alpha$ of the simulation costs $2^{\alpha t}$, where $t$ is the number of $T$ gates in a RQC.
    Blue, orange, and green lines correspond to the scaling using stabilizer-state sampling, optimized stabilizer-state sampling, and Heisenberg propagation, respectively.
  }
  \label{fig:sim_cost_scaling}
\end{figure}
We also investigate the scaling factor $\alpha$ of the simulation costs $2^{\alpha t}$ with various error rates of the single-qubit and two-qubit depolarizing noise $p$ (\cref{fig:sim_cost_scaling}).
The blue, orange, and green lines correspond to the scaling for stabilizer-state sampling, optimized stabilizer-state sampling, and Heisenberg propagation, respectively.
The sampling cost of error-free RQCs for the optimized stabilizer-state sampling is the lowest, while that for Heisenberg propagation is the highest.
In this case, the scaling based on Heisenberg propagation is $2^t$, which is the same as the scaling based on stabilizer-state sampling when the resource state is decomposed over only separable stabilizer states.
Furthermore, when the error rate is high, the simulation costs of noisy quantum circuits for stabilizer-state sampling and Heisenberg propagation are lower than the simulation costs for the stabilizer rank algorithm $2^{0.468 t}$, ~\cite{bravyiImprovedClassicalSimulation2016,bravyiTradingClassicalQuantum2016}, which is a well-known sampling-based simulator for pure states.
Specifically, scaling based on Heisenberg propagation and optimized stabilizer-state sampling is lower than scaling based on the stabilizer rank when $p\geq 0.1$ and $p\geq 0.13$, respectively.
Additionally, \cref{fig:sim_cost_scaling} confirms that the threshold error rate of classical simulatability of Heisenberg propagation is lower than that of stabilizer-state sampling.

Finally, we consider the time required to calculate these noisy RQCs.
We assume that we can estimate the expectation value within an additive error $\delta = 10^{-3}$, with a success probability of at least $1 - \epsilon = 1 - 10^{-2}$ and the error rate of the single-qubit and two-qubit depolarizing noise $p=0.05$.
For $p=0.05$, $\alpha=0.66$ and $\alpha=0.73$ for optimized stabilizer-state sampling and Heisenberg propagation, respectively.
Combining the above information, we find that the simulation of the noisy random circuits with $t=40$ requires approximately $8.9 \times 10^{14}$ samples for the optimized stabilizer-state sampling and $6.2 \times 10^{15}$ samples for Heisenberg propagation.
We also assume that we have a classical computer with $10^6$ CPU cores, and each CPU core takes 1 ms to calculate one sample, which is what a typical implementation of this algorithm takes.
In this case, the computer needs about 10 s using optimized stabilizer-state sampling and 70 s using Heisenberg propagation.

\section{Conclusions and discussion}
We evaluate the simulation costs of noisy quantum circuits by two sampling-based classical algorithms: stabilizer-state sampling and Heisenberg propagation.
We extend and improve the existing stabilizer-state sampling algorithm for noisy quantum circuits.
To compare the two sampling-based classical simulation algorithms, we also investigate the simulation costs of noisy RQCs.
We find that for a low error rate, stabilizer-state sampling is the better algorithm, otherwise, Heisenberg propagation is the better.
Scaling based on the simulation costs of noisy RQCs via Heisenberg propagation and optimized stabilizer-state sampling is lower than that by the stabilizer-rank simulator~\cite{bravyiImprovedClassicalSimulation2016,bravyiTradingClassicalQuantum2016} 
when $p\leq 0.1$ and $p\leq 0.13$, respectively.

Recently, Seddon \textit{et al}.~\cite{seddonQuantifyingQuantumSpeedups2021} presented classical sampling algorithms and their associated magic monotones,
where a resource state is decomposed,
instead of using density operators of stabilizer states, 
into a matrix spanned by pure stabilizer states, so-called stabilizer dyads.
Since they showed that the dyadic frame simulator is faster than stabilizer-state sampling,
it will be interesting to apply a dyadic frame simulator 
for noisy quantum circuits.

We believe that the knowledge obtained regarding the sampling-based 
classical simulation methods is useful not only for the classical simulation itself but also for designing applications of near-term quantum devices.
Specifically,
one of the most-promising applications of near-term quantum devices, VQE, uses parameterized quantum circuits, which consist of many rotation gates.
In particular, the angles of the rotation gates of the unitary coupled-cluster ansatz with the Trotter decomposition are typically very small.
As discussed in Section~\ref{sec:sampling_costs_of_noisy_circuits}, a small-angle rotation gate can be easily simulated by stabilizer-state sampling.
Even if a unitary coupled-cluster ansatz without Trotterization or a hardware-efficient ansatz is used, the angles would be small.
This is because a Hartree-Fock state as the initial state of the VQE
is a good approximation of the ground state.
In Ref.~\cite{mitaraiQuadraticCliffordExpansion2020}, a perturbative approach is sufficient to describe the VQE with a hardware-efficient ansatz, implying the rotations are very small.
Furthermore, for the unitary coupled-cluster ansatz, many two-qubit Clifford gates are used to construct the ansatz.
Noise on two-qubit Clifford gates deteriorates the ROM of the rotation gates, which would make the parameterized circuit classically simulatable more easily.
With these points in mind, we have to design the parameterized quantum circuits under noise carefully so that the VQE has a potential quantum advantage.
It would be interesting to characterize the Hamiltonian and the ansatz of the VQE with respect to the ROM of the quantum circuit generating an optimal variational solution.

\begin{acknowledgments}
  K.\ F.\  is supported by KAKENHI Grant No.\ 16H02211, JST PRESTO JPMJPR1668, JST ERATO JPMJER1601, and JST CREST JPMJCR1673. This work is supported by MEXT, Q-LEAP Grants No.\  JPMXS0118067394 and No.\ JPMXS0120319794. 
  This paper is partially based on results obtained from a project commissioned by the New Energy and Industrial Technology Development Organization (NEDO). 
\end{acknowledgments}

\bibliography{main}

\end{document}